\documentclass[journal,twocolumn]{IEEEtran}
\usepackage{amsfonts, amsmath, algorithm,algpseudocode,subfigure, color,cases,bm,amssymb}
\usepackage{graphicx,epstopdf}

\newtheorem{proposition}{Proposition}[section]

\DeclareMathAlphabet{\mathpzc}{OT1}{pzc}{m}{it}
\newcommand{\comment}[1]{}
\def\Re{\text{Re}}
\def\Im{\text{Im}}

\def\zero{\boldsymbol{0}}

\def\bbe{\boldsymbol{\beta}}

\def\bLa{\boldsymbol{\Lambda}}

\def\({\left(}
\def\){\right)}
\def\[{\left[}
\def\]{\right]}
\def\BEq{\begin{eqnarray}}
\def\EEq{\end{eqnarray}}
\def\BE*{\begin{eqnarray*}}
\def\EE*{\end{eqnarray*}}
\def\BA{\begin{array}}
\def\EA{\end{array}}

\def\0{\mathbf{0}}
\def\1{\mathbf{1}}
\def\a{\mathbf{a}}
\def\A{\mathbf{A}}

\def\bC{\mathbb{C}}

\def\D{\mathbf{D}}

\def\bE{\mathbb{E}}

\def\F{\mathbf{F}}

\def\g{\mathbf{g}}
\def\G{\mathbf{G}}

\def\I{\mathbf{I}}

\def\n{\mathbf{n}}
\def\N{\mathbf{N}}

\def\R{\mathbf{R}}

\def\bR{\mathbb{R}}

\def\bfu{\mathbf{u}}

\def\bfv{\mathbf{v}}

\def\W{\mathbf{W}}
\def\x{\mathbf{x}}
\def\X{\mathbf{X}}
\def\y{\mathbf{y}}

\def\z{\mathbf{z}}

\def\and{\prefixtext{and}}

\def\diag{{\rm diag}}

\newcommand{\sgn}{\mathrm{sgn}}
\newcommand{\pr}{\mathrm{Pr}}
\allowdisplaybreaks
\graphicspath{{Figures/}}

\newtheorem{Lemma}{Lemma}[section]
\newtheorem{Col}{Corollary}[section]

\newtheorem{Remark}{Remark}[section]

\begin{document}

\title{Inexact Alternating Optimization for Phase Retrieval In the Presence of Outliers}

\author{Cheng Qian, Xiao Fu, \emph{Member, IEEE}, Nicholas D. Sidiropoulos, \emph{Fellow, IEEE}, Lei Huang, \emph{Senior Member, IEEE}, and Junhao Xie, \emph{Senior Member, IEEE}
\thanks{
Original manuscript submitted to {\em IEEE Trans. on Signal Processing}, October 4, 2016; revised \today. Conference
version of part of this work appears in {\em Proc. EUSIPCO 2016} \cite{EUSIPCO2016}. \par
C. Qian, X. Fu and N. D. Sidiropoulos are with the Department of Electrical and Computer Engineering, University of Minnesota, Minneapolis, MN 55455 USA (e-mail: alextoqc@gmail.com, xfu@umn.edu, nikos@umn.edu).\par
L. Huang is with the College of Information Engineering, Shenzhen University, Shenzhen, 518060 China (e-mail: dr.lei.huang@ieee.org). \par
J. Xie is with the Department of Electronics and Information Engineering, Harbin Institute of Technology, Harbin, 150001 China (e-mail: xj@hit.edu.cn). \par
The work of N. Sidiropoulos was supported by NSF CIF-1525194.
}
}
\maketitle

\begin{abstract}
Phase retrieval has been mainly considered in the presence of Gaussian noise. However, the performance of the algorithms proposed under the Gaussian noise model severely degrades when grossly corrupted data, i.e., outliers, exist. This paper investigates techniques for phase retrieval in the presence of heavy-tailed noise -- which is considered a better model for situations where outliers exist. An $\ell_p$-norm ($0<p<2$) based estimator is proposed for fending against such noise, and two-block inexact alternating optimization is proposed as the algorithmic framework to tackle the resulting optimization problem. Two specific algorithms are devised by exploring different local approximations within this framework. Interestingly, the core conditional minimization steps can be interpreted as iteratively reweighted least squares and gradient descent. Convergence properties of the algorithms are discussed, and the Cram\'er-Rao bound (CRB) is derived. Simulations demonstrate that the proposed algorithms approach the CRB and outperform state-of-the-art algorithms in heavy-tailed noise.
\end{abstract}


\section{Introduction}
\textit{Phase retrieval} aims at recovering a signal $\x\in\mathbb{C}^{N}$ from only the magnitude of linear measurements.
This is an old problem \cite{GS,Fienup} that has recently attracted renewed and growing interest. Phase retrieval arises in many fields, such as X-ray crystallography, coherent diffraction imaging, and optical imaging and astronomy, where the detectors only record the intensity information, because phase is very difficult and expensive to measure.

The mathematical description of the phase retrieval problem is simple: given the measuring matrix ${\bf A}=[\a_1\ \cdots\ \a_M]^H\in\mathbb{C}^{M\times N}$ and the measurement vector ${\bf y}\in\mathbb{R}^M$, where $\a_m\in\bC^N$, and
\begin{equation*}
	{\bf y}=|{\bf A}{\bf x}|,
\end{equation*}
find ${\bf x} \in\mathbb{C}^{N}$.
Early attempts to solve the phase retrieval problem can be traced back to the 1970s, where techniques such as Gerchberg-Saxton (GS) \cite{GS}, Fienup \cite{Fienup} and their variants were proposed. These algorithms have been empirically shown to work well under certain conditions, although little had been known regarding their convergence properties from a theoretical point of view. Recently, some new algorithms along this line of work were proposed in \cite{AltMinPr}, where the convergence issue is better studied. 

In recent years, more modern optimization-based approaches for phase retrieval have been proposed.
For example, Cand\`es proposed a semidefinite relaxation approach known as the \emph{PhaseLift} algorithm \cite{C1} and proved that exact recovery is possible with high probability in the noiseless case.
Hand recently studied the robustness of PhaseLift and showed that it can tolerate a constant fraction of arbitrary errors \cite{robustPL}.
Wirtinger-Flow (WF) \cite{WF} is a more recent approach that combines a good statistical initialization with a computationally light gradient-type refinement algorithm. The combination works very well when ${\bf A}$ is i.i.d. Gaussian.
Following \cite{WF}, conceptually similar approaches, namely, the truncated WF (TWF) \cite{trunctedWF} and truncated amplitude flow (TAF) algorithms \cite{TAF,STAF} were proposed to handle more challenging scenarios.
In the recent work in \cite{FPP_PR,FPP_PR2}, the authors proposed a least-squares {\em feasible point pursuit} (LS-FPP) approach that aims to solve the same optimization problem as LS PhaseLift in the presence of noise. Simulations in \cite{FPP_PR} indicate that LS-FPP approaches the Cram\'er-Rao bound (CRB) for the Gaussian measurement model considered in \cite{FPP_PR}.

Some phase retrieval algorithms were originally developed under an exact measurement model, and subsequently treated the noisy case by replacing equality constraints with relaxed inequality constraints \cite{C1}-\cite{C2}. Most of the existing algorithms were explicitly or implicitly developed under a Gaussian noise model. In certain applications, a subset of the measurements may be corrupted much more significantly than the others, and heavy-tailed noise may be encountered as well \cite{IRLS_Quadratic}-\cite{ImpulsiveNoise}. One representative example is high energy coherent X-ray imaging using a charge-coupled device (CCD), where the impulsive noise originates from X-ray radiation effects on the CCD, and the density of impulses increases with the intensity of X-ray radiation or CCD exposure time \cite{ImpulsiveNoise}-\cite{implusivenoise1}. Under such circumstance, modeling the noise as Gaussian is no longer appropriate.

In recent years, robust phase retrieval algorithms, e.g., \cite{IRLS_Quadratic,PR_Spar_IN2,MTWF}, have been proposed to handle outliers.
The framework in \cite{PR_Spar_IN2} considered an undersampled phase retrieval model corrupted with Laplacian-distributed outliers,
but the approach was designed specifically for sparse ${\bf x}$.
	Variations of TWF to handle outliers were also considered.
	For example, the work in  \cite{MTWF} replaces the sample mean that is used in TWF by sample median in the initialization and truncated update, which exhibits robustness to outliers under certain conditions \cite{MTWF}. However, as TWF, this approach still heavily relies on the assumption that Gaussian measurements are employed.

%

Another important aspect of phase retrieval is how noise enters the measurement model. Our previous work \cite{FPP_PR} considered the noise model $\y = |\A\x|^2 + \n$. Another noise model that is frequently considered in the literature \cite{GS}-\cite{AltMinPr} is
\begin{align}\label{model}
  \y = |\A\x| + \n.
\end{align}
Notice the subtle but important difference between the two models: whether noise is added to the magnitude or the squared magnitude. The choice hinges on the experimental setup, including the measurement apparatus; but \eqref{model} is more widely adopted by experimentalists.

\noindent
{\bf Contributions}:
We consider the phase retrieval problem under the model in \eqref{model} in the presence of impulsive noise, and focus on designing robust algorithms to handle it.
To fend against impulsive noise, we adopt the $\ell_p$-fitting ($0<p<2$) based estimator that is known to be effective in dealing with outliers, and devise two optimization algorithms using two-block inexact alternating optimization.
Specifically, the two algorithms both solve local majorization-based approximate subproblems for one block, instead of exactly solving the conditional block minimization problem to optimality.
Interestingly, starting from different majorizations, the resulting solutions turn out equivalent to iteratively reweighted least squares and gradient descent, respectively.
Unlike the existing inexact and exact alternating optimization frameworks that mostly operate with convex constraints,
the proposed algorithms work with a unit-modulus nonconvex constraint, so convergence analysis seems difficult.
Nevertheless, we prove convergence of the proposed algorithms to a Karush-Kuhn-Tucker (KKT) point by exploiting the two-block structure of the problem.
We also derive computationally light implementations using Nesterov-type and stochastic gradient updates.
In order to assist in performance analysis and experimental design, we derive the CRB for the model in \eqref{model} and under different parameterizations, in the presence of Laplacian and Gaussian noise.
Curiously, although related CRBs for different noisy measurement models have been previously derived in \cite{FPP_PR}, \cite{crb_pr0}-\cite{crb_pr4}, to the best of our knowledge, there is no available CRB for the model in \eqref{model} -- and our work fills this gap.
The proposed algorithms are validated by extensive simulations.
The simulations show that our approaches outperform the state-of-the-art algorithms in the presence of impulsive noise.

A preliminary conference version of part of this work has been submitted to EUSIPCO 2016 \cite{EUSIPCO2016}. The conference version includes one of the two basic algorithms, the Laplacian CRB, and limited simulations. The second algorithm, Nesterov acceleration and stochastic gradient-type updates, additional CRB results, and, most importantly, proof of convergence of the iterative algorithms are all provided only in this journal version, which naturally also includes more comprehensive simulation results.

\noindent
{\bf Notation:} Throughout the paper, we use boldface lowercase letters for vectors and boldface uppercase letters for matrices. Superscripts $(\cdot)^T$, $(\cdot)^*$, $(\cdot)^H$, $(\cdot)^{-1}$ and $(\cdot)^\dagger$ represent transpose, complex conjugate, conjugate transpose, matrix inverse and pseudo-inverse, respectively. The $\Re\{\cdot\}$ and $\Im\{\cdot\}$ denote the real part and imaginary part. $\bE[\cdot]$ is the expectation operator, $|\cdot|$ is the absolute value operator, $||\cdot||_F$ is the Frobenius norm, $||\cdot||_p$ is the vector $\ell_p$-norm, whose definition is $||\x||_p = (\sum_{i=1}^N|x_i|^p)^{1/p}$, $\odot$ is the element-wise product, and $\text{diag}(\cdot)$ is a diagonal matrix with its argument on the diagonal. $\delta_{ij}$ denotes the Kronecker delta function, and $\angle(\cdot)$ takes the phase of its argument. $\boldsymbol{0}_m$, $\boldsymbol 1_m$, and $\I_m$ stand for the $m\times 1$ all-zero vector, $m\times 1$ all-one vector, and $m\times m$ identity matrix, respectively. Furthermore, $\text{trace}(\cdot)$ and $\partial a/\partial x$ denote the trace and partial derivative operators, respectively.


\section{Proposed Algorithms}
\subsection{AltIRLS}
Let us first consider the noiseless case where ${\bf y}=|{\bf A}{\bf x}|$.
Effectively, it can also be written as
\begin{equation}\label{eq:eff_model}
{\bf y}\odot{\bf u}={\bf A}{\bf x}
\end{equation}
where ${\bf u}= e^{j\angle(\A {\bf x})}$ is an auxiliary vector of unknown unit-modulus variables with its $m$th component being $u_m = e^{j\angle(\a_m^H\x)}$. In the presence of impulsive noise, $\ell_p$-(quasi)-norm has be recognized as an effective tool for promoting sparsity and fending against outliers \cite{volmin}-\cite{outlier4}.
Therefore, we propose to employ an $\ell_p$-fitting based estimator instead of using the $\ell_2$-norm, which has the form of
\begin{align}\label{P00}
	\min_{|\bfu|=\boldsymbol{1},\x} \sum_{m=1}^{M}\left(|y_mu_m - \a_m^H\x |^2 + \epsilon\right)^{p/2}
\end{align}
where $0<p<2$ is chosen to down-weigh noise impulses (i.e., outliers), and
$\epsilon>0$ is a small regularization parameter (e.g., $\epsilon\in[10^{-8},10^{-6}]$) that keeps the cost function within its differentiable domain, which will prove handy in devising an effective algorithm later.

To handle Problem~\eqref{P00}, we follow the rationale of alternating optimization, i.e.,
we first update ${\bf x}$ fixing ${\bf u}$, and then we do the same for ${\bf u}$.

Assume that after some iterations, the current solution at iteration $r$ is $(\x^{(r)}, \bfu^{(r)})$. At step $(r+1)$, the subproblem with respect to (w.r.t.) ${\bf x}$ is
\begin{align}\label{P1}
{\bf x}^{(r+1)}=\arg	\min_{\x} \sum_{m=1}^{M}\left(|y_mu_m^{(r)} - \a_m^H\x |^2 + \epsilon\right)^{p/2}
\end{align}
which is still difficult to handle.
Particularly, when $p<1$, the subproblem itself is still non-convex;
when $p\geq 1$, the subproblem is convex but has no closed-form solution.
Under such circumstances, we propose to employ the following lemma \cite{xf}:
\begin{Lemma}\label{lem:conjugate}
     Assume $0<p<2$, $\epsilon> 0$, and
		$\phi_p(w) := \frac{2-p}{2}\left(\frac{2}{p}w \right)^{\frac{p}{p-2}} +\epsilon w$.
		Then, we have
		\begin{equation}
		\left(x^2+\epsilon\right)^{p/2} = \min_{w\geq 0}~w x^2+\phi_p(w)
		\end{equation}
		and the unique minimizer is
		\begin{equation}
            w_{\rm opt} = \frac{p}{2}\left(x^2+\epsilon\right)^{\frac{p-2}{2}}.
		\end{equation}
\end{Lemma}
By Lemma \ref{lem:conjugate}, an upper bound of $\sum_{m=1}^{M}\left(|y_mu_m^{(r)} - \a_m^H\x |^2 + \epsilon\right)^{p/2}$ that is tight at the current solution ${\bf x}^{(r)}$ can be easily found:
\begin{align}\label{eq:P1_upp}
	 &\sum_{m=1}^{M}\left(|y_mu_m^{(r)} - \a_m^H\x |^2 + \epsilon\right)^{p/2} \notag\\ &\qquad\leq \sum_{m=1}^M \left(w_m^{(r)}\left|y_m u_m^{(r)}-{\bf a}^H_{m}{\bf x}\right|^2 +\phi_p\left(w_m^{(r)}\right)\right)
\end{align}
where
\begin{align}\label{eq:w_def}
  w_m^{(r)}:=\frac{p}{2}\left(\left|y_m u_m^{(r)} - \a_m^H{\x}^{(r)}\right|^2 + \epsilon\right)^{\frac{p-2}{2}}
\end{align}
and the equality holds if and only if ${\bf x}={\bf x}^{(r)}$.
Instead of directly dealing with Problem~\eqref{P1},
we solve a surrogate problem using the right hand side (RHS) of \eqref{eq:P1_upp} at each iteration to update ${\bf x}$.
Notice that the RHS of \eqref{eq:P1_upp} is convex w.r.t. ${\bf x}$ and the corresponding problem can be solved in closed-form:
\begin{equation}\label{wls}
   \x^{(r+1)} = \big(\W^{(r)}\A\big)^\dagger\W^{(r)}\big(\y\odot{\bf u}^{(r)}\big)
\end{equation}
where
\begin{align}\label{W}
  {\bf W}^{(r)}={\rm diag}\left(\sqrt{w^{(r)}_1}\ \cdots\ \sqrt{w^{(r)}_M}\right).
\end{align}

The conditional problem w.r.t. ${\bf u}$ is
\begin{align}\label{P0}
	{\bf u}^{(r+1)}=\arg\min_{|\bfu|=\boldsymbol{1}} \sum_{m=1}^{M}\left(|y_mu_m - \a_m^H\x^{(r+1)} |^2 + \epsilon\right)^{p/2}.
\end{align}
Although the problem is non-convex, it can be easily solved to optimality.
Specifically,
the first observation is that the partial minimization w.r.t. ${\bf u}$ is insensitive to the value of $p$; i.e., given a fixed ${\bf x}$,
for any $p>0$, the solutions w.r.t. ${\bf u}$ are identical.
Second, for all $p>0$, the solution is simply to align the angle of $y_mu_m$ to that of ${\bf a}^H_m{\bf x}^{(r+1)}$,
which is exactly
\begin{align}\label{uk1}
	{u}_m^{(r+1)}= e^{j \angle\left(\a_m^H {\bf x}^{(r+1)}\right)},\quad m=1,\ldots,M.
\end{align}

We update ${\bf x}$ and ${\bf u}$ alternately, until some convergence criterion is met.
We see that the way that we construct the upper bound of the partial problem w.r.t. ${\bf x}$ is in fact the same
as the procedure in iteratively reweighted least squares (IRLS) \cite{outlier4,xf}. The difference is that we `embed' this IRLS
step into an alternating optimization algorithm.
We therefore call this algorithm alternating IRLS (AltIRLS), which is summarized in Algorithm~\ref{Alg_IRLS}.

\begin{algorithm}[ht]
\caption{AltIRLS for phase retrieval}
\begin{algorithmic}[1]
\Function{AltIRLS  }{$\y,\A,\x^{(0)}$}
\State Initialize $\bfu^{(0)}=\exp(\angle(\A\x^{(0)}))$ and $\W^{(0)}={\bf W}^{(0)}={\rm diag}\left(\sqrt{w^{(0)}_1}\ \cdots\ \sqrt{w^{(0)}_M}\right)$ with $w^{(0)}_m = \frac{p}{2}\left(\left|y_m u_m^{(0)} - \a_m^H{\x}^{(0)}\right|^2 + \epsilon\right)^{\frac{p-2}{2}}$
\While{stopping criterion has not been reached}
\State $\x^{(r)} = (\W^{(r-1)}\A)^\dagger\W^{(r-1)}(\y\odot{\bf u}^{(r-1)})$.
\State $\bfu^{(r)} = \exp(j\angle(\A\x^{(r)}))$
\State $w_m^{(r)}=\frac{p}{2}\left(\left|y_m u_m^{(r)} - \a_m^H{\x}^{(r)}\right|^2 + \epsilon\right)^{\frac{p-2}{2}},\forall m$
\State $\W^{(r)} = {\rm diag}\left(\sqrt{w^{(r)}_1}\ \cdots\ \sqrt{w^{(r)}_M}\right)$
\State $r=r+1$
\EndWhile
\EndFunction
\end{algorithmic}\label{Alg_IRLS}
\end{algorithm}

A relevant question regarding Algorithm~\ref{Alg_IRLS} is whether or not this algorithm converges to a meaningful point, e.g., a stationary or KKT point.
Note that the block variable ${\bf u}$ is constrained to a non-convex set, and we do not compute the optimal solution for the block variable ${\bf x}$ at each iteration of Algorithm~\ref{Alg_IRLS}.
For such a type of algorithm, there is no existing theoretical framework that establishes convergence.
We therefore need careful custom convergence analysis. We have the following result:
\begin{proposition}\label{prop:convergence}
	Assume that $0<p<2$ and $\epsilon>0$, and that $\A$ has full column rank.
	Then, the solution sequence produced by Algorithm~\ref{Alg_IRLS} converges to a set ${\cal K}$ that consists of all the KKT points of Problem~\eqref{P00}. 
\end{proposition}
\begin{IEEEproof}
  See Appendix \ref{Appendix_convergence}.
\end{IEEEproof}
The result in Proposition~\ref{prop:convergence} is interesting: Although the algorithm that we propose to compute the $\ell_p$ fitting-based estimator solves non-convex subproblems, it ensures convergence to a KKT point.
Such a nice property is proven by exploiting the two-block structure of the algorithm.
Note that the proof itself does not rely on the specific form of the optimization problem in \eqref{P00}, and thus can be easily extended to other two-block alternating optimization cases, which we believe is of much broader interest.

\subsection{AltGD}
Algorithm~\ref{Alg_IRLS} uses a simple update strategy, but its complexity may still become an issue when the problem size grows.
Specifically, the bottleneck of the AltIRLS algorithm lies in solving the subproblem w.r.t. $\x$, i.e.,
\begin{align}\label{minx}
 \min_\x\ \left|\left| \W^{(r)}(\y\odot\bfu) - \W^{(r)}\A\x \right|\right|_2^2
\end{align}
whose closed-form solution can be computed via \eqref{wls}. It is observed from \eqref{wls} that
the matrix inversion part requires $\mathcal O(N^3)$ flops to compute and also $\mathcal O(N^2)$ memory to store, both of which are not efficient for high-dimensional $\x$ -- e.g., when $\x$ is a vectorized image, $N$ is usually very large (more specifically, for a $100\times100$ image, $N$ is 10,000).

To circumvent this difficulty, one way is to employ a certain fast least squares solver such as conjugate gradient \cite{cg} to handle Problem~\eqref{minx}. This is a viable option, but such solvers still typically require many iterations to obtain a fairly accurate solution of \eqref{minx}.
Here, we propose to deal with problem \eqref{minx} using a simpler approach. Specifically, let us denote
\begin{align}
  f^{(r)}(\x) =\left|\left| \W^{(r)}(\y\odot\bfu^{(r)}) - \W^{(r)}\A\x \right|\right|_2^2
\end{align}
and approximate \eqref{minx} using the following function
\begin{align}\label{mm}
  g^{(r)}(\x)& = f^{(r)}(\x^{(r)}) +{\rm Re}\{(\nabla f^{(r)}(\x^{(r)}))^H(\x-\x^{(r)})\} \nonumber\\
	             &\quad\quad+ \frac{\mu^{(r)}}{2}\left\|\x-\x^{(r)}\right\|_2^2.
\end{align}
where $\mu^{(r)}\geq0$ is a pre-specified parameter.
Instead of optimizing $f^{(r)}({\bf x})$, we optimize $\g^{(r)}(\x)$. By rearranging terms, the subproblem becomes
\begin{align}
  \x^{(r+1)} =\arg\min_\x\ \left\|\x - \left( \x^{(r)} - \frac{1}{\mu^{(r)}}\nabla f(\x^{(r)}) \right)\right\|_2^2
\end{align}
and the solution is
\begin{align}
  \x^{(r+1)} = \x^{(r)} - \frac{1}{\mu^{(r)}}\nabla f(\x^{(r)})
\end{align}
i.e., a gradient step with step-size $1/\mu^{(r)}$, where the gradient is
\begin{align}
  \nabla f^{(r)}(\x^{(r)}) = \A^H\big(\W^{(r)}\big)^2\big(\A\x^{(r)} - \y\odot\bfu^{(r)}\big)
\end{align}
which does not require any matrix inversion operation. Also, the $N\times N$ matrix does not need to be stored, if we take the order $\A\x^{(r)}\rightarrow\big(\W^{(r)}\big)^2\A\x^{(r)}\rightarrow\A^H\big(\W^{(r)}\big)^2\A\x^{(r)}$ to compute the update of ${\bf x}$.
Therefore, using this approach, both memory and complexity requirements are less demanding.
We summarize the algorithm in Algorithm~\ref{algo:GD}. We call this algorithm alternating gradient descent (AltGD).

Note that when $\mu$ is chosen as the Lipschitz constant, \eqref{mm} is an upper bound of the cost function in \eqref{minx}.
In other words, we have
\begin{equation*}
	 f^{(r)}({\bf x})\leq g^{(r)}({\bf x})
\end{equation*}
if $\mu^{(r)}\geq \lambda_{\max}({\bf A}^H({\bf W}^{(r)})^2{\bf A})$,
and the equality holds if and only if ${\bf x}={\bf x}^{(r)}$.
Therefore, the algorithmic structure of Algorithm~\ref{algo:GD} is the same as that of Algorithm~\ref{Alg_IRLS},
except that the majorizing functions of the ${\bf x}$-block are different -- which means that the proof of convergence of Algorithm~\ref{Alg_IRLS}
also applies here:
\begin{Col}
	If $\mu^{(r)}\geq \lambda_{\max}({\bf A}^T({\bf W}^{(r)})^2{\bf A})$ for all $r$,
    then the whole sequence produced by Algorithm~\ref{Alg_IRLS} converges to ${\cal K}$.
\end{Col}

\begin{algorithm}[ht]
\caption{AltGD for phase retrieval}\label{algo:GD}
\label{CHalgorithm}
\begin{algorithmic}[1]
\Function{AltGD}{$\y,\A,\x^{(0)}$}
\State Initialize $\bfu^{(0)}=\exp(\angle(\A\x^{(0)}))$
\While{stopping criterion has not been reached}
\State Choose $\mu^{(r-1)}$ as the leading eigenvalue of $\A^H(\W^{(r-1)})^2\A$ or the trace of $(\W^{(r-1)})^2$
\State $\x^{(r)} = \x^{(r-1)} - 1/\mu^{(r-1)}\nabla f(\x^{(r-1)})$
\State $\bfu^{(r)} = \exp(j\angle(\A\x^{(r)}))$
\State $r=r+1$
\EndWhile
\EndFunction
\end{algorithmic}
\end{algorithm}

\begin{Remark}\label{remark}
Exactly computing $\lambda_{\max}({\bf A}^H({\bf W}^{(r)})^2{\bf A})$ may be time consuming in practice.
Many practically easier ways can be employed, e.g., the Armijo rule-based line search \cite{beck2009fast}.
In our simulations, we use a simple heuristic: we let $\mu^{(r)} =\text{trace}((\W^{(r)})^2)$ instead of $\lambda_{\max}({\bf A}^H({\bf W}^{(r)})^2{\bf A})$.
The rationale behind is that we observe that the energy of ${\bf A}^H({\bf W}^{(r)})^2{\bf A}$ is usually dominated by ${\bf W}^{(r)}$, and using
$\mu^{(r)} =\text{trace}((\W^{(r)})^2)$ is a good approximation of ${\rm trace}({\bf A}^H({\bf W}^{(r)})^2{\bf A})$ that is an upper bound of $\lambda_{\max}({\bf A}^H({\bf W}^{(r)})^2{\bf A})$. We should mention that this step-size
choice is a heuristic, but works well in practice, as will be shown in the simulations.

\end{Remark}

\subsection{Further Reducing Complexity}
\subsubsection{Extrapolation}
\begin{figure}
  \includegraphics[width=8cm]{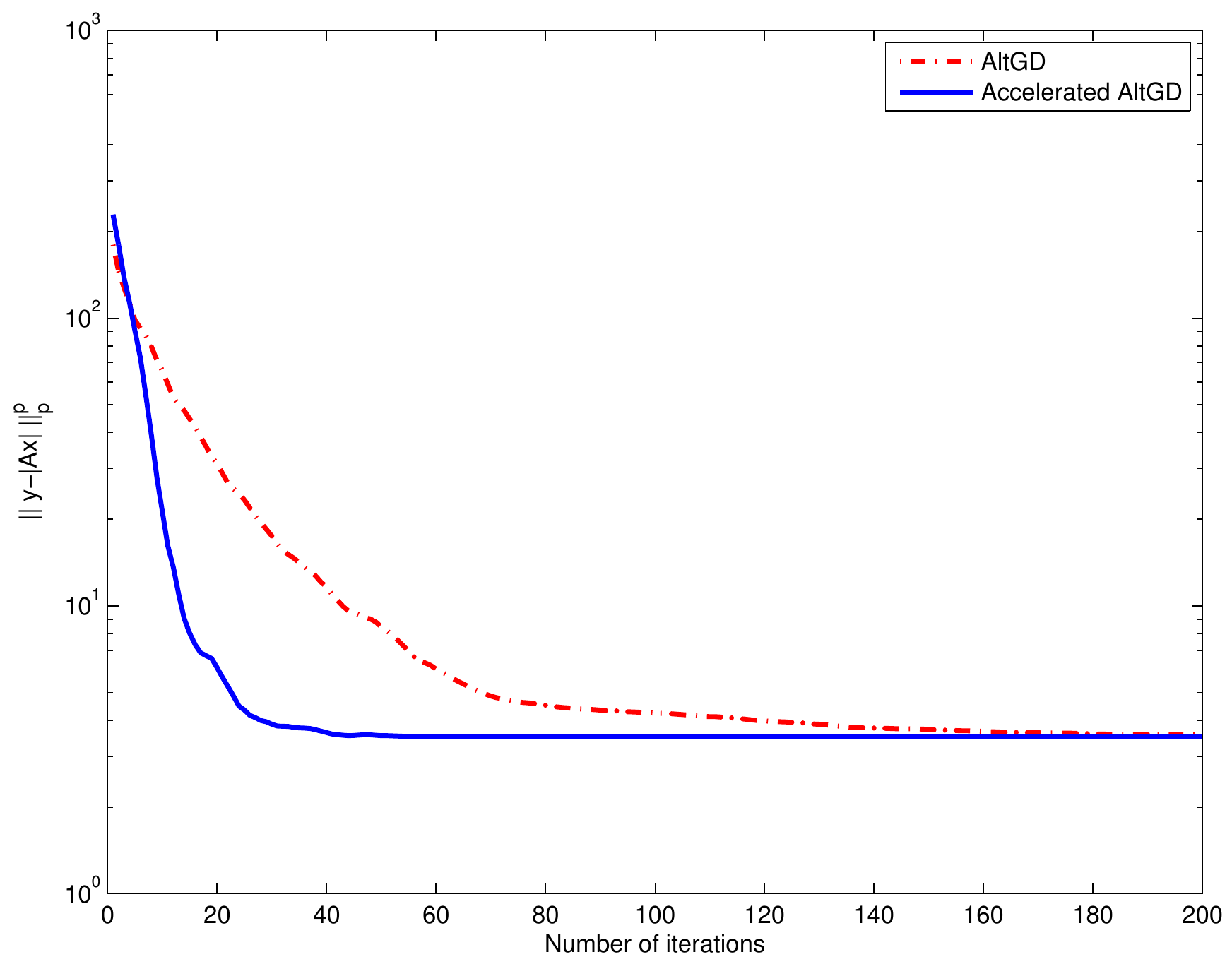}\\
  \caption{Cost function value versus number of iterations.}\label{CostFuncValue}
\end{figure}

Algorithm 2 is easier to compute than Algorithm 1 in terms of per-iteration complexity. However, first-order methods tend to require more iterations in total. One way to alleviate this effect is to incorporate Nesterov's ``optimal first-order'' method \cite{nesterov1}-\cite{nesterov3}, i.e., for each update, we set
\begin{align}
  \z^{(r)} =&\ \x^{(r)} + \frac{t^{(r-1)}-1}{t^{(r)}}\left(\x^{(r)}-\x^{(r-1)}\right) \\
  t^{(r)} =&\ \frac{1+\sqrt{1+4(t^{(r-1)})^2}}{2} \\
  \x^{(r+1)} =&\ \z^{(r)} - \frac{1}{\mu^{(r)}}\nabla f^{(r)}(\z^{(r)})
\end{align}
In practice, the above `extrapolation' technique greatly expedites the whole process in various applications \cite{volmin,xuyin}.
Some numerical evidence can be seen in Fig.~\ref{CostFuncValue}, where a simple comparison between the plain Algorithm~2 and the extrapolated version is presented.
We choose SNR$=20$ dB, $N=16$ and $M=128$. Each element in the signal and measurement vectors is independently drawn from the complex circularly symmetric Gaussian distribution with mean zero and variance one. The noise is generated from a symmetric $\alpha$ stable (S$\alpha$S distribution) which will be described in detail in Section IV.  Fig. \ref{CostFuncValue} shows the convergence of AltGD and accelerated AltGD using extrapolation when $p=1.3$. As we can see, the accelerated AltGD converges after 40 iterations which is much faster than AltGD that converges after 200 iterations.

\subsubsection{Block Incremental / Stochastic Gradient}
When $N$ is very large (thus $M>N$ is larger), even gradient computation is too much of a burden. In such cases, a pragmatic way is to ``break down'' the problem to pieces and do (block) incremental or stochastic gradient.
We divide the measurement matrix into $L$ blocks $\Gamma_l$.
Then it is straightforward to get the gradient for the $l$th block as
\begin{align}\label{g2}
  \nabla f_{\Gamma_l}(\x^{(r)}) = \A^H_{\Gamma_l}\big(\W_{\Gamma_l}^{(r)}\big)^2\big(\A_{\Gamma_l}\x^{(r)} - \y_{\Gamma_l}\odot\bfu_{\Gamma_l}^{(r)}\big).
\end{align}
The estimate of $\x$ is updated according to
\begin{align}
  \x^{(r+1)} = \x^{(r)} - \frac{1}{\mu^{(r)}_{\Gamma_l}}\nabla f_{\Gamma_l}(\x^{(r)})
\end{align}
where $\mu^{(r)}_{\Gamma_l}$ is chosen as the leading eigenvalue of $\A_{\Gamma_l}^H\A_{\Gamma_l}$ or ${\rm trace}\big(\big(\W_{\Gamma_l}^{(r)}\big)^2\big)$.
The algorithm can proceed block by block with revisits, or by randomly picking blocks, resulting in block incremental gradient and stochastic gradient versions, respectively.


A subtle point here is that, to maintain robustness, one should choose a block size larger than one.
The reason is that the robustness of the algorithm is brought by treating different $y_m$ with different weights (more specifically, by downweighting the outliers). When using only one $y_m$ for updating, this ability vanishes.



\section{Cram\'er-Rao Bound Analysis}
In this section, the CRB on the accuracy of retrieving $\x$ in \eqref{model} is presented. The CRB provides a lower bound on the MSE of unbiased estimators. 
In many signal processing applications, one cannot guarantee that an estimator is unbiased, or even that an unbiased estimator exists, yet the CRB is still surprisingly predictive of estimator performance \cite{crb4,basu2000}.
For phase retrieval under Gaussian noise, over the past few years, several CRBs have been derived for different models, e.g., 2-D Fourier-based measurements \cite{crb_pr0}, noise added prior to taking the magnitude \cite{crb_pr1} and noise added after taking the magnitude square \cite{FPP_PR}, \cite{crb_pr3}. To the best of our knowledge, there is no available CRB formula for the signal model in \eqref{model}. Here, we present the CRBs for two particular types of noise: Laplacian and Gaussian noise. Although
our main interest here lies in evaluating performance of robust algorithms and the Laplacian CRB serves this purpose, the Gaussian CRB is of interest in other application contexts. Note that we use subscripts \emph{r, c, L} and \emph{G} to stand for real, complex, Laplacian and Gaussian, respectively.

To get started, we should note that our derivations are based on the assumption that ${\bf a}_m^H{\bf x}$ is nonzero. That is because the term $|\A\x|$ is non-differentiable at ${\bf a}_m^H{\bf x} = 0$, the Fisher information only exists at ${\bf a}_m^H{\bf x}\neq 0$.
With this caveat, we have the following proposition:
\begin{proposition}\label{proposition1}
  In Laplacian noise, the CRB is
  \begin{align}\label{CRBc}
	\mathrm{CRB}_{L,c} = \mathrm{trace}\left(\F_{L,c}^\dagger \right)
  \end{align}
where
\begin{align}
  \F_{L,c} = \frac{2}{\sigma_n^2}\G_{L,c}\,\diag(|\A\x|^{-2})\,\G^T_{L,c}
\end{align}
with
\begin{align}
	\G_{L,c} =
	\begin{bmatrix}
	  \Re\{\A^H\diag(\A\x)\}\\
	  \Im\{\A^H\diag(\A\x)\}
	\end{bmatrix}.
\end{align}
\end{proposition}
\begin{IEEEproof}
  See Appendix \ref{Appendix_CRB_Laplacian}.
\end{IEEEproof}
The CRB for real $\x$ is a special case of the complex case, which can be easily derived from Proposition \ref{proposition1}:
\begin{proposition}\label{proposition12}
  In Laplacian noise, the CRB of real-valued $\x$ is
  \begin{align}\label{CRBr}
	\mathrm{CRB}_{L,r} = \mathrm{trace}\left(\F_{L,r}^{-1} \right)
  \end{align}
where
\begin{align}\label{laplacianFIM}
  \F_{L,r} = \frac{2}{\sigma_n^2}\G_{L,r}\,\diag(|\A\x|^{-2})\,\G_{L,r}^T
\end{align}
with
\begin{align}
	\G_{L,r} = \Re\{\A^H\diag(\A\x)\}.
\end{align}
\end{proposition}

In Appendix \ref{Appendix_F_rank}, we show that when $\A$ has full column rank, $\F_{L,c}$ is singular with rank $(2N-1)$ while $\F_{L,r}$ is always nonsingular. Thus, for complex $\x$, we adopt its pseudo-inverse to compute an optimistic (looser) CRB, which is still a valid lower bound that can be used to benchmark the efficiency of any biased estimator \cite{crb1}-\cite{crb4}. If $\x$ is close to zero, then the CRB is not tight at low SNRs, and more measurements should be used to approach the CRB.
\begin{proposition}\label{proposition2}
  In the Gaussian noise case, the CRB is two times larger than the CRB in Proposition \ref{proposition1}, i.e.,
  \begin{align}\label{crb_gaussian}
    \mathrm{CRB}_{G} = 2\mathrm{CRB}_{L}.
  \end{align}
\end{proposition}
\begin{IEEEproof}
	The proof is straightforward by calculating the Gaussian FIM and comparing it with the Laplacian FIM in \eqref{laplacianFIM}. We omit it here.
\end{IEEEproof}

In certain applications of phase retrieval, we may be more interested in the performance bound for retrieving the phase of $\x$. The following proposition provides the CRB on both the phase and amplitude of $\x$ under Laplacian noise. Note that in the Gaussian noise case, it is straightforward to apply Proposition \ref{proposition2} to compute the lower bound. Similar results can also be found in \cite{LapNoiseCRB}.
\begin{proposition}\label{proposition3}
  In Laplacian noise, the CRB of the amplitude of $\x$ is
  \begin{align}\label{CRB_amplitude}
	\mathrm{CRB}_{L,|\x|} = \sum_{i=1}^N d_i
  \end{align}
  and the variance of any {biased} estimate of the phase of $\x$ is bounded below by
  \begin{align}\label{CRB_phase}
	\mathrm{CRB}_{L,\angle(\x)} = \sum_{i=N+1}^{2N} d_i
  \end{align}
where $d = [d_1\ \cdots\ d_{2N}]$ contains the main diagonal elements of $\F_L^\dagger$ which in this case is defined as
\begin{align}
    \F_L = \frac{2}{\sigma_n^2}\G_L\,\diag(|\A\x|^{-2})\,\G^T_L
\end{align}
with
\begin{align}
  \G_L =
  \begin{bmatrix}
    \diag(|\x|)^{-1} \!\!&\!\! \\ &\!\! \I_N
  \end{bmatrix}\!\!
	\begin{bmatrix}
	  \Re\left\{\diag\left(\x^*\right)\A^H\diag(\A\x)\right\}\\
	  \Im\{\diag(\x^*)\A^H\diag(\A\x)\}
	\end{bmatrix}.
\end{align}
\end{proposition}
\begin{IEEEproof}
	See \cite{FPP_PR} and Appendix \ref{Appendix_CRB_Laplacian}.
\end{IEEEproof}
\begin{Remark}
  In deriving the CRB in Proposition \ref{proposition3}, we use no additional assumptions on $\x$. Therefore, Proposition \ref{proposition3} works for both real and complex $\x$. Specifically, in the real case, the phase is actually the sign of $\x$. Here, $\F_L$ is also singular with rank deficit equal to one, so we adopt its pseudo-inverse to compute the CRB. We omit the proof of rank-1 deficiency of $\F_L$, since it follows the line of argument in Appendix \ref{Appendix_F_rank}.
\end{Remark}


\section{Simulation Results}
In this section, we evaluate the performance of the proposed methods and compare them with some existing algorithms, namely, WF \cite{WF}, TWF \cite{trunctedWF}, TAF \cite{TAF}, AltMinPhase \cite{AltMinPr}, MTWF \cite{MTWF} and GS \cite{GS} in terms of MSE performance, where the MSE is computed after removing the global phase ambiguity between the true and estimated $\x$. In the simulations, we consider an exponential signal $\x = \exp(j0.16\pi t)$ comprising $16$ samples, i.e., $t=1,\cdots,16$.
We test the algorithms using different types of ${\bf A}$, i.e., random matrix that is usually employed in \cite{AltMinPr}, \cite{WF}-\cite{TAF,MTWF}, and the 2D Fourier matrix that is widely used in real devices \cite{GS,Fienup} (cf. subsection \ref{sec:2dfourier}).
For the random measuring matrix case (used in subsections A-D), the measurement vectors are generated from a masked Fourier transformation, which takes the form of
\begin{equation*}
  \A =
    \begin{bmatrix}
       (\D\bLa_1)^T & \cdots & (\D\bLa_{K})^T
    \end{bmatrix}^T
\end{equation*}
where $K=M/N$ is the number of masks, $\D$ is a $N\times N$ discrete Fourier transform (DFT) matrix with $\D\D^H=N\I_N$ and $\bLa_k$ is a $N\times N$ diagonal masking matrix with its diagonal entries generated by $b_1b_2$, where $b_1$ and $b_2$ are independent and distributed as \cite{C1}
\begin{equation*}
   b_1 = \left\{
   \begin{aligned}
      1\quad  &\text{w. prob. 0.25} \\
      -1\quad &\text{w. prob. 0.25} \\
      -j\quad &\text{w. prob. 0.25} \\
      j\quad  &\text{w. prob. 0.25}
   \end{aligned} \right.
\ \text{and}\
b_2 = \left\{
      \begin{aligned}
         \sqrt{2}/2\quad  &\text{w. prob. 0.8} \\
         \sqrt{3}\quad &\text{w. prob. 0.2}.
      \end{aligned} \right.
\end{equation*}
All results are obtained using a computer with 3.1 GHz Intel Xeon E31225  and 16 GB RAM. For the random measurement case, the algorithms are all initialized from the same starting point that is computed by picking the principal eigenvector of $\sum_{i=1}^My_i^2\a_i\a_i^H$, and the stopping criterion is
$$\frac{\Big| \left|\left|y-|\A\x^{(r)}|\right|\right|_2^2-\left|\left|y-|\A\x^{(r-1)}| \right|\right|_2^2 \Big|}{\left|\left|y-|\A\x^{(r-1)}|\right|\right|_2^2}\leq10^{-7}$$ or the number of iterations reaching 1000.
Furthermore, we use AltGD with extrapolation for the simulations.

\subsection{Selection of $p$}
\begin{figure}
  \centering
  \includegraphics[width=1\linewidth]{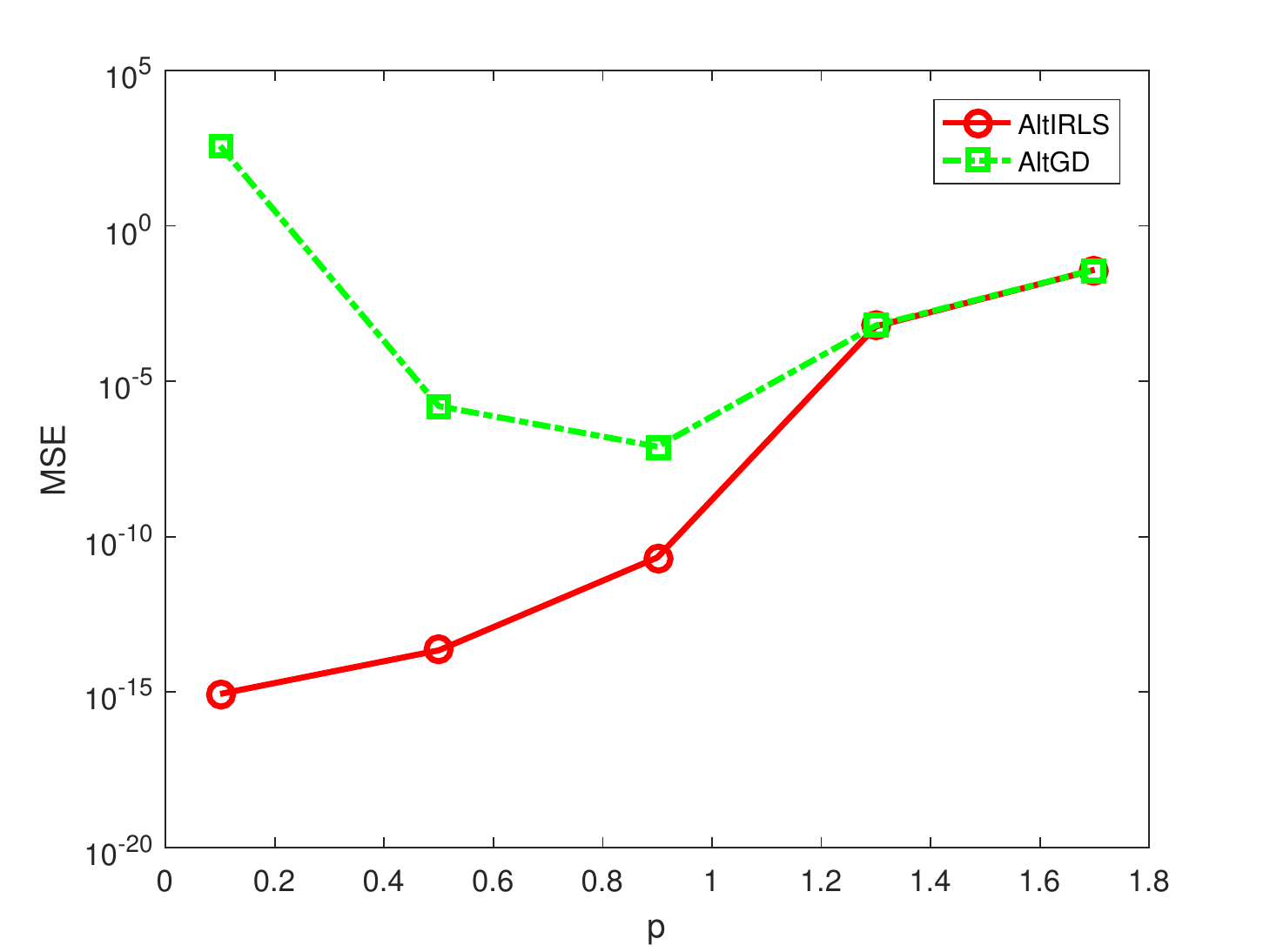}\\
  \caption{MSE versus $p$.}\label{fig:p}
\end{figure}

Before we do the performance comparison, let us study how $p$ affects the performance of the proposed methods. In this example, SNR is fixed at 20 dB, where the SNR is computed via
$$\text{SNR} = 10\log_{10}\left(\frac{||\A\x||^2}{||\n||^2}\right).$$
We assume that 10\% of the data are corrupted by outliers that are generated from the Gaussian distribution with mean zero and variance 100, and the remaining elements in $\n$ are zero. $K=8$ masks are employed to generate the measurements. Fig. \ref{fig:p} shows the MSE as a function of $p$. Note that, when $p$ is smaller than 1, in order to achieve convergence in a non-convex setting, we initialize our methods in a two-step way. Specifically, for each of the proposed methods, when $0.6< p<1$, we use the spectrum output to initialize our methods with $p=1.3$ and 100 iterations, and then choose the corresponding output to do another 100 iterations with $p=1$ and use the output as a final starting point. When $p\leq 0.6$, we take an additional intermediate step with $p=0.7$ and 100 iterations to gradually stage the initialization process for our approaches, since in this $p$-regime the subproblem is strictly non-convex. It is observed that generally, $p\leq1$ provides better performance than $p>1$, especially for AltIRLS. When $p<0.5$, AltGD suffers severe performance degradation while AltIRLS still performs well. Our understanding is that first-order algorithms are in general more sensitive to problem conditioning, and a small $p$ can lead to badly-conditioned ${\bf W}^{(r)}$. Working with $p<1$ usually requires much more careful initialization. In our experience, $p \approx 1.3$ strikes a good compromise between numerical stability and estimation accuracy.


\begin{figure}
  \centering
  \includegraphics[width=1\linewidth]{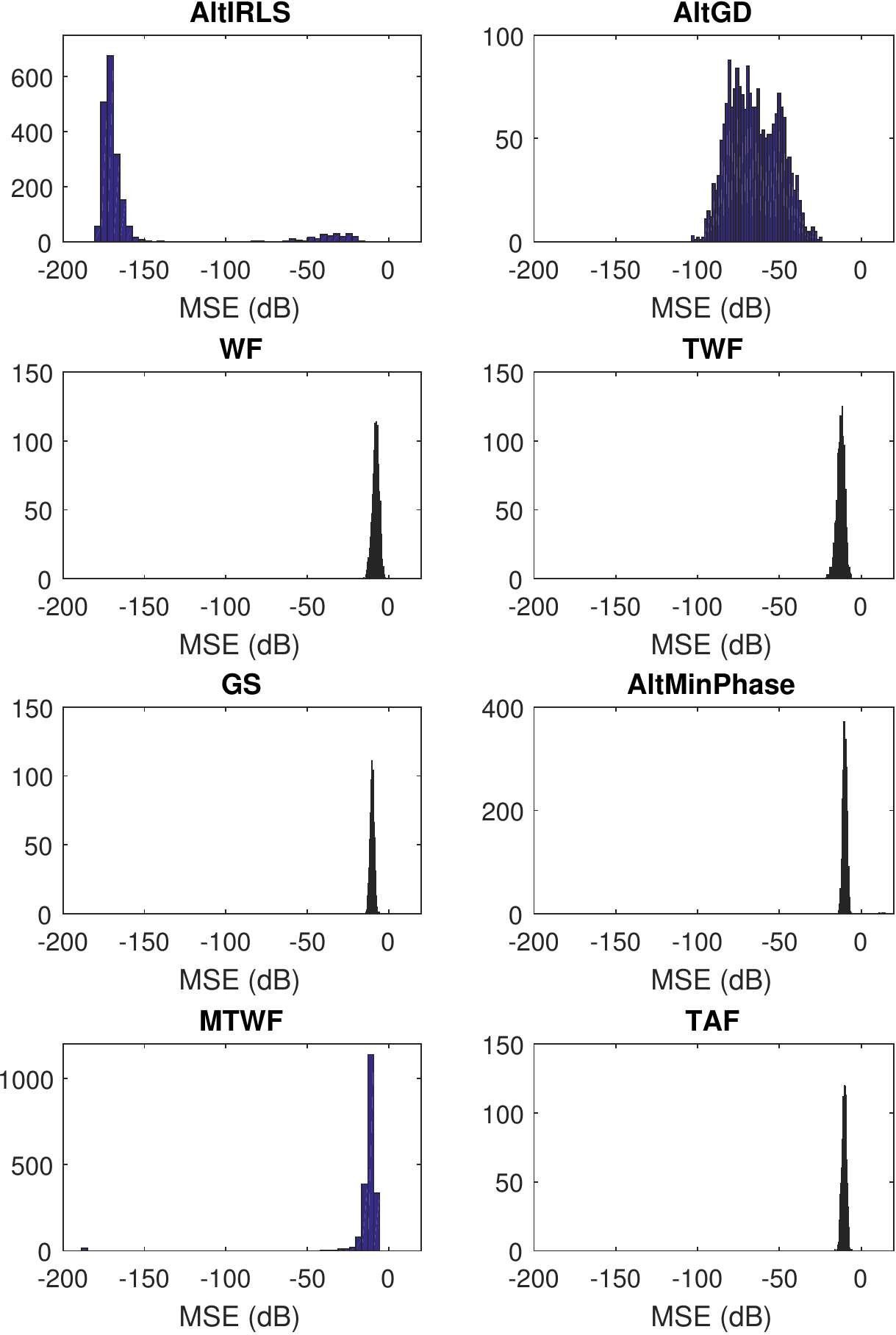}\\
  \caption{Signal recovery performance comparison.}\label{hist}
\end{figure}

To illustrate the performance provided by our proposed methods with $p<1$ over the state-of-art algorithms including TWF, WF, TAF, AltMinPhase, MTWF, and GS, we present the following example, where the parameters are the same as for Fig. \ref{fig:p}, except that 20\% of the data are corrupted by outliers.
Note that among these competitors, TAF, AltMinPhase and GS share the same magnitude measurement model in \eqref{model} as ours, but WF is designed under the energy measurement model (i.e., ${y}_m=|{\bf a}_m^H{\bf x}|^2 + n_m$). In addition, TWF and MTWF adopt a Poisson model with $\lambda=|\a_i^H\x|^2$.
Hence, $\y^2$ is fed to WF, TWF and MTWF, where the squaring is element-wise.
We set $p=0.4$ for the proposed methods, and perform Monte-Carlo trials to estimate $\text{MSE}=10\log_{10}(||\hat\x-\x||_2^2)$.
We show the results of 1,000 random trials in Fig.~\ref{hist}.
One can see that for most trials the benchmark algorithms fail to give reasonable results.
However, our methods produce much lower MSEs, which indicates that our proposed methods are efficient in suppressing outliers.

\subsection{Statistical Performance Comparison}
We now compare the MSE performance as a function of SNR. The results are averaged over 500 Monte-Carlo trials.
To simulate outliers, we use the Laplacian distribution, the $\alpha$-stable distribution, and a Gaussian mixture model (GMM) to generate heavy-tailed noise ${\bf n}$, respectively. Specifically, the outlier generating process is summarized below:

\subsubsection{Laplacian}
The probability density function (PDF) of the Laplacian distribution is given in \eqref{pdf_lap}.

\subsubsection{$\alpha$-stable}
The PDF of an $\alpha$-stable distribution is generally not available, but its characteristic function can be written in closed-form as
$$
  \phi(t; \alpha,\beta,c,\mu) = \exp\left( jt\mu - \gamma^\alpha|t|^\alpha\left(1-j\beta\mathrm{sgn}(t)\Phi(\alpha)\right) \right)
$$
where $\Phi(\alpha)=\tan\left(\alpha\pi/2\right)$, $0<\alpha\leq2$ is the stability parameter, $-1\leq\beta\leq1$ is a measure of asymmetry, $\gamma>0$ is a scale factor which measures the width of the distribution and $\mu$ is a shift parameter. There are two special cases that admit closed-form PDF expression, that is, $\alpha=1$ and $\alpha=2$ which respectively yield the Cauchy and Gaussian distributions. For $\alpha<2$, $\phi(t; \alpha,\beta,\gamma,\mu)$ possesses heavy tails, and thus is considered suitable for simulating impulsive noise. The parameter $\alpha$ controls the density of impulses - smaller $\alpha$'s correspond to more outliers. In the following, we set $\alpha=0.8$, and the other parameters are $\beta=0$, $\gamma=2$ and $\mu=0$, resulting in a symmetric $\alpha$-stable (S$\alpha$S) distribution with zero-shift.

\subsubsection{GMM}
In our simulations,
we also use a two-component GMM to generate impulsive noise, whose PDF is as follows:
\begin{align}
p(n) =  \sum_{i=1}^2 \frac{c_i}{\sqrt{2\pi\sigma_i^2}}\exp\left\{ -\frac{|n|^2}{2\sigma_i^2} \right\}
\end{align}
where $\sigma_i^2$ is the variance of the $i$th term, and $0\leq c_i\leq 1$ is the probability of occurrence of the $i$th component. Note that we have $c_1+c_2=1$, and we set $c_2<c_1$ and $\sigma_2^2>\sigma_1^2$ -- therefore, the second component corresponds to outliers. In this example, we choose $c_1=0.9$, $c_2=0.1$, $\sigma_1^2=0.1$ and $\sigma_2^2=100$, which corresponds to a situation where strong outliers are present.

~\smallskip

\begin{figure}
    \begin{center}
        \includegraphics[width=1\linewidth]{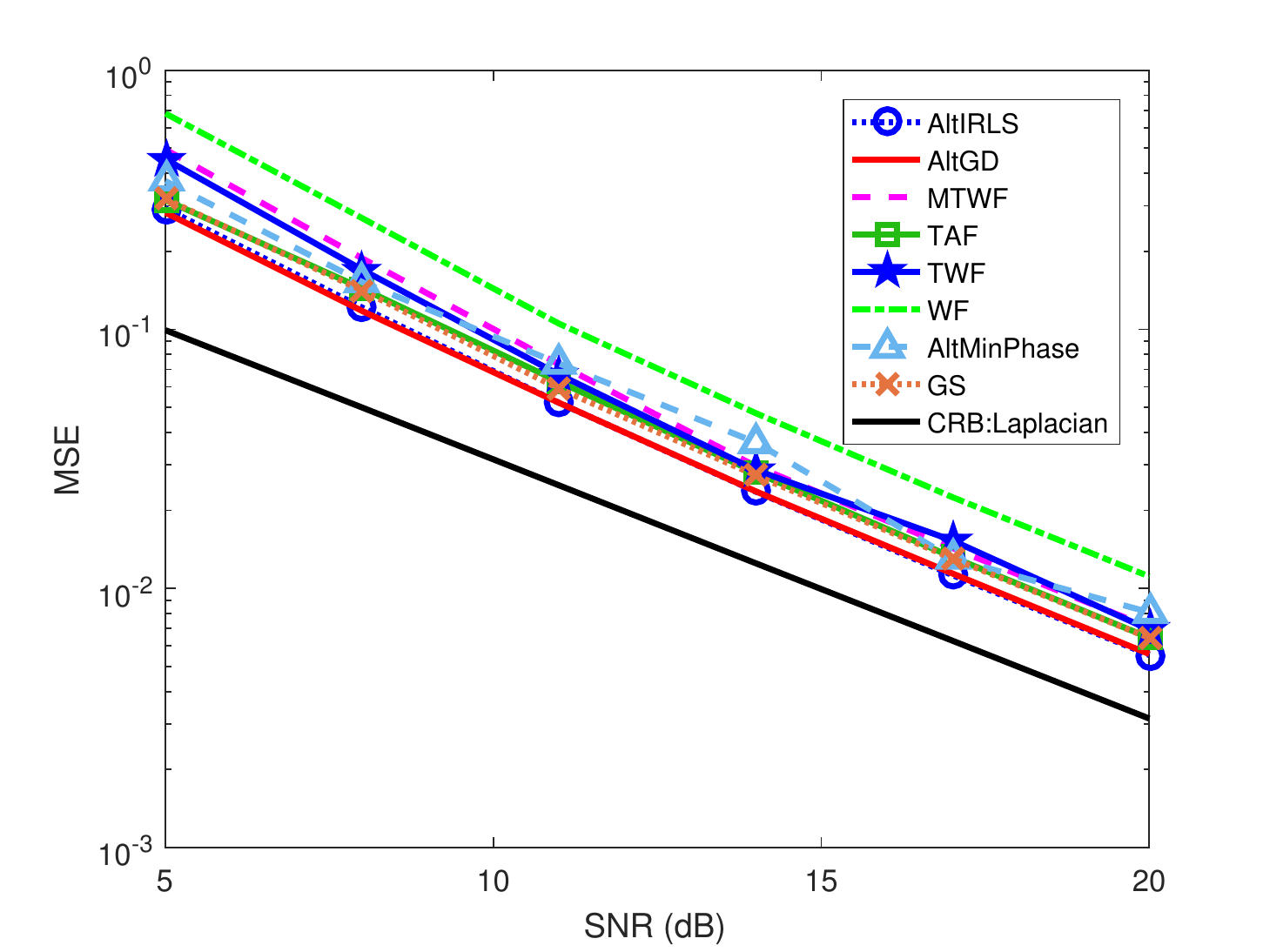}
        \caption{MSE versus SNR in Laplacian noise.}\label{Fig_L}
    \end{center}
\end{figure}

\begin{figure}
    \begin{center}
        \includegraphics[width=1\linewidth]{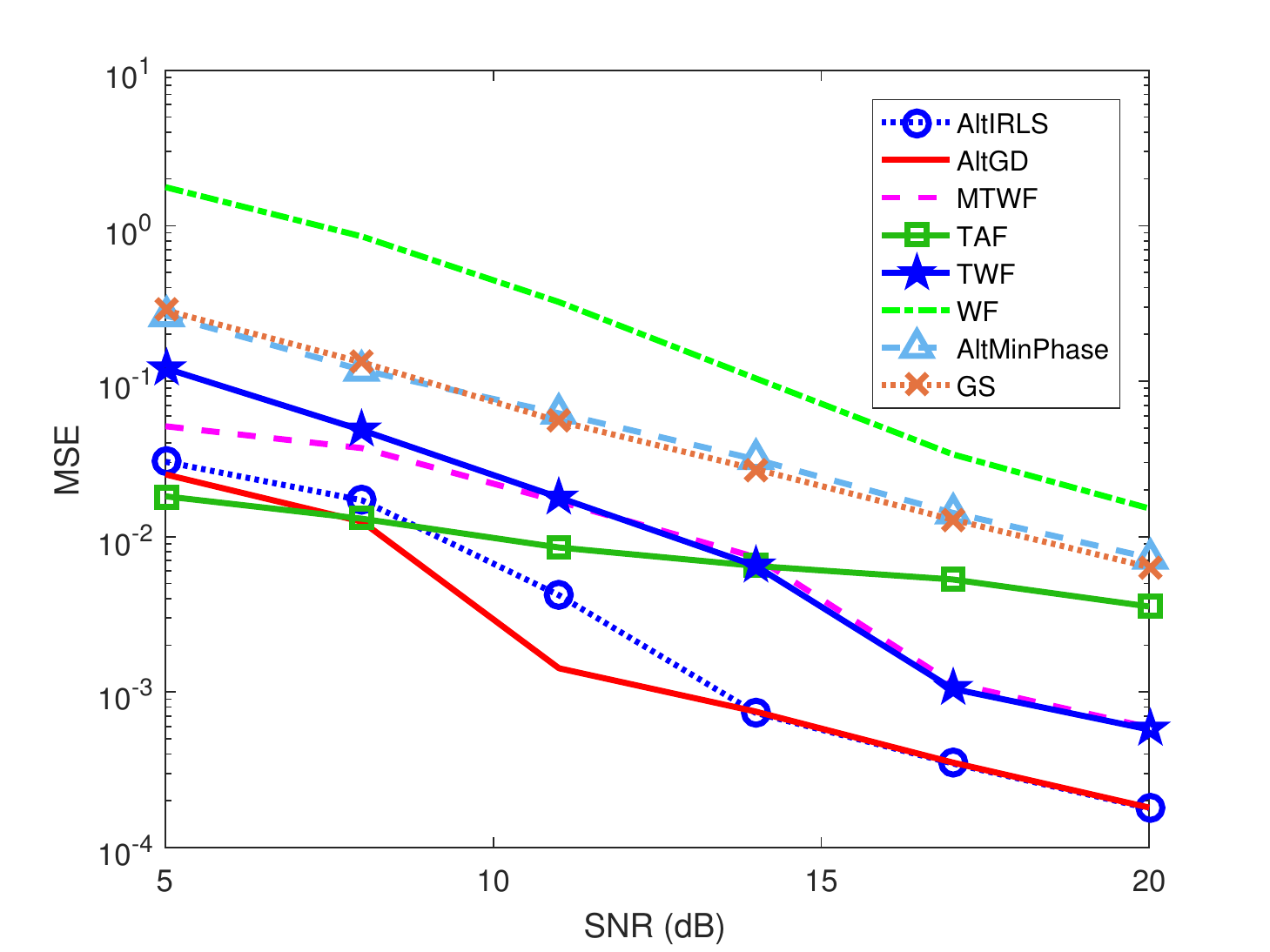}
        \caption{MSE versus SNR in S$\alpha$S noise.}\label{Fig_AS}
    \end{center}
\end{figure}

\begin{figure}
    \begin{center}
        \includegraphics[width=1\linewidth]{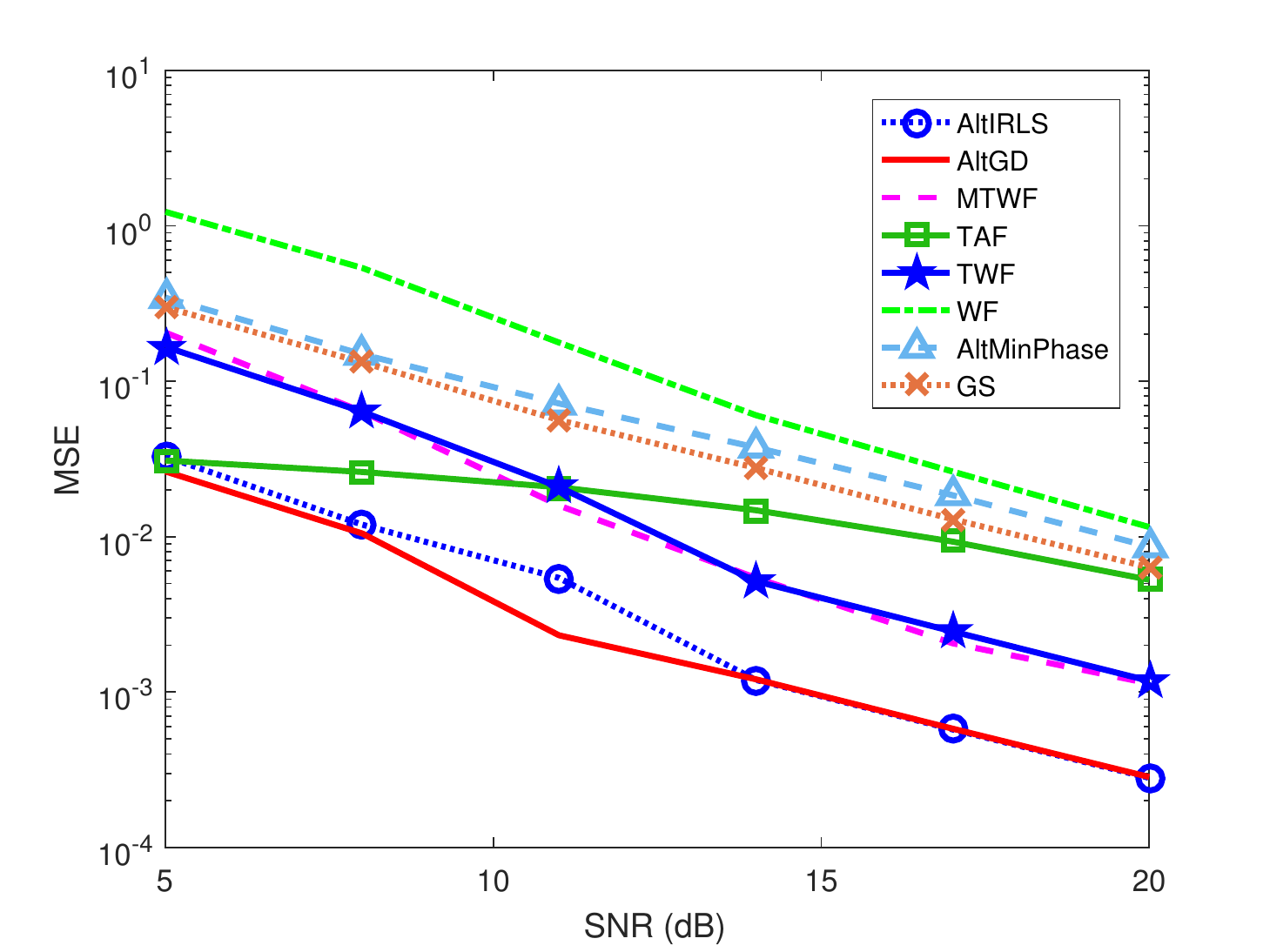}
        \caption{MSE versus SNR in GMM noise.}\label{Fig_GMM}
    \end{center}
\end{figure}

Figs. \ref{Fig_L}, \ref{Fig_AS} and \ref{Fig_GMM} show the MSEs of $\hat\x$ under Laplacian, S$\alpha$S and GMM noise, respectively. In the simulations, the number of measurements is set to be $M=8N$.
Note that for S$\alpha$S and GMM noise, we set $p=1.3$ for AltIRLS and AltGD, while for Laplacian noise we let $p=1$ which corresponds to the maximum likelihood estimator.
In the Laplacian noise case, we include the CRB derived in Proposition \ref{proposition12} as a performance benchmark.
It is observed from Fig. \ref{Fig_L} that our approaches outperform the GS algorithm and yield the MSEs which are closest to the CRB.
The performance gap between the proposed methods and their competitors becomes larger in the S$\alpha$S and GMM noise cases since the outliers become stronger; see Fig. \ref{Fig_AS} and \ref{Fig_GMM}.
This indicates that the proposed algorithms work better in more critical situations, i.e., when more severe outliers exist. Note that the MTWF performs slightly better than TWF in GMM noise and its performance is inferior to our methods in the three noise scenarios.

\begin{figure}
    \begin{center}
        \includegraphics[width=1\linewidth]{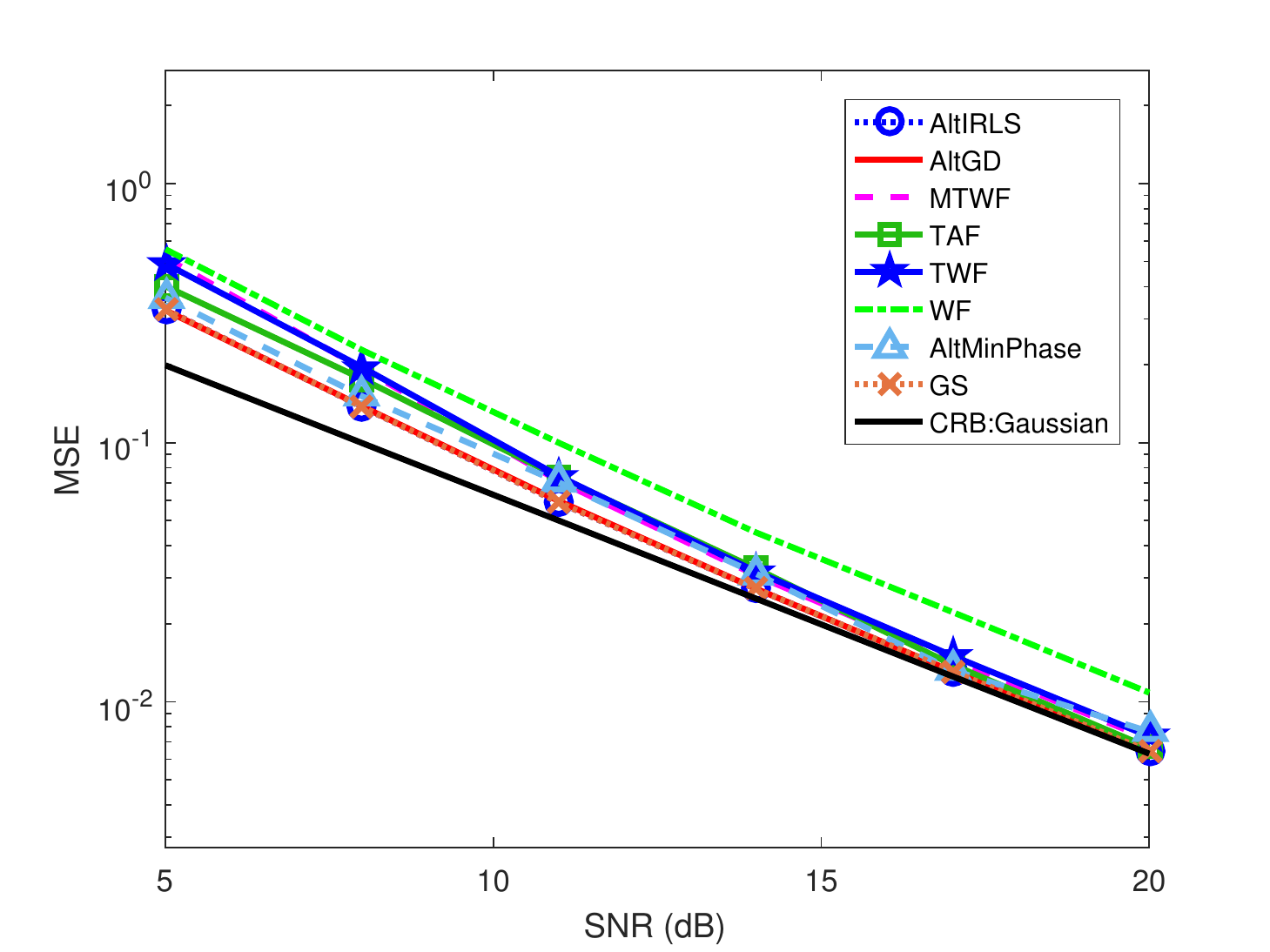}
        \caption{MSE versus SNR in Gaussian noise.}\label{Fig_GN}
    \end{center}
\end{figure}

As a reference, we also plot the MSE performance versus SNR in Gaussian noise in Fig. \ref{Fig_GN}, where we set $p=2$ for having a maximum likelihood estimator.
We compare the performance of the algorithms with the CRB in Proposition~\ref{proposition3}.
One can see that the algorithms perform similarly when the noise follows an i.i.d. Gaussian distribution. This suggests that the Gaussian noise case may be an easier case - where all the algorithms under test perform reasonably.

\begin{figure}
    \begin{center}
        \includegraphics[width=1\linewidth]{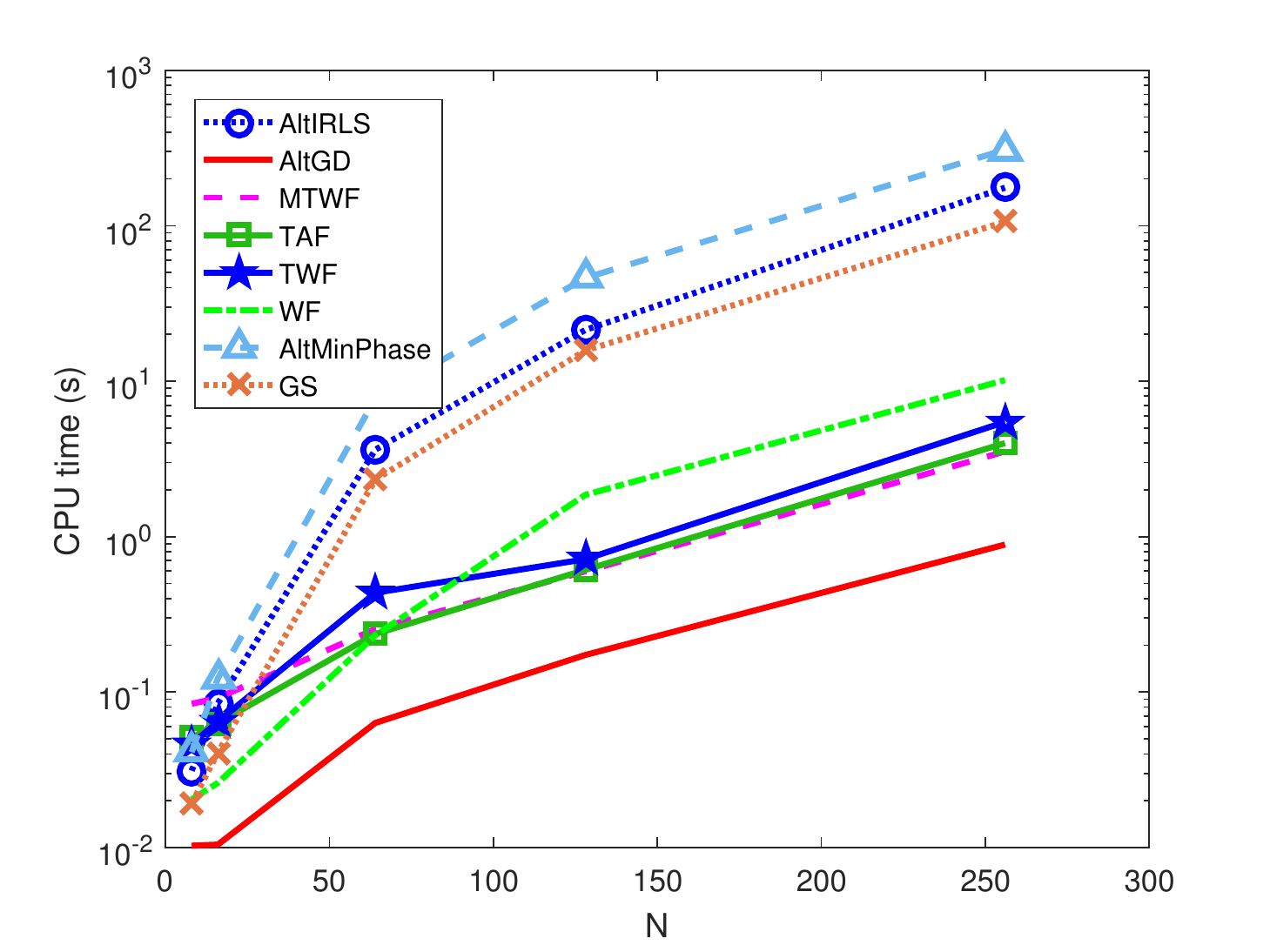}
        \caption{CPU time versus number of variables $N$ in GMM noise.}\label{Fig_time}
    \end{center}
\end{figure}

Fig. \ref{Fig_time} compares the CPU time as a function of $N$, where SNR is 10 dB, $N$ increases from 8 to 128 and the other parameters are the same as Fig. \ref{Fig_GMM}. It is seen that AltGD is the fastest\footnote{Note that in the simulation we use the heuristic step size as stated in Remark \ref{remark}; if the step size $\mu=\lambda_{\max}({\bf A}^H({\bf W}^{(r)})^2{\bf A})$ is used, AltGD may consume more time.}.

\subsection{Success Rate Performance Comparison}
\begin{figure}
    \begin{center}
        \includegraphics[width=1\linewidth]{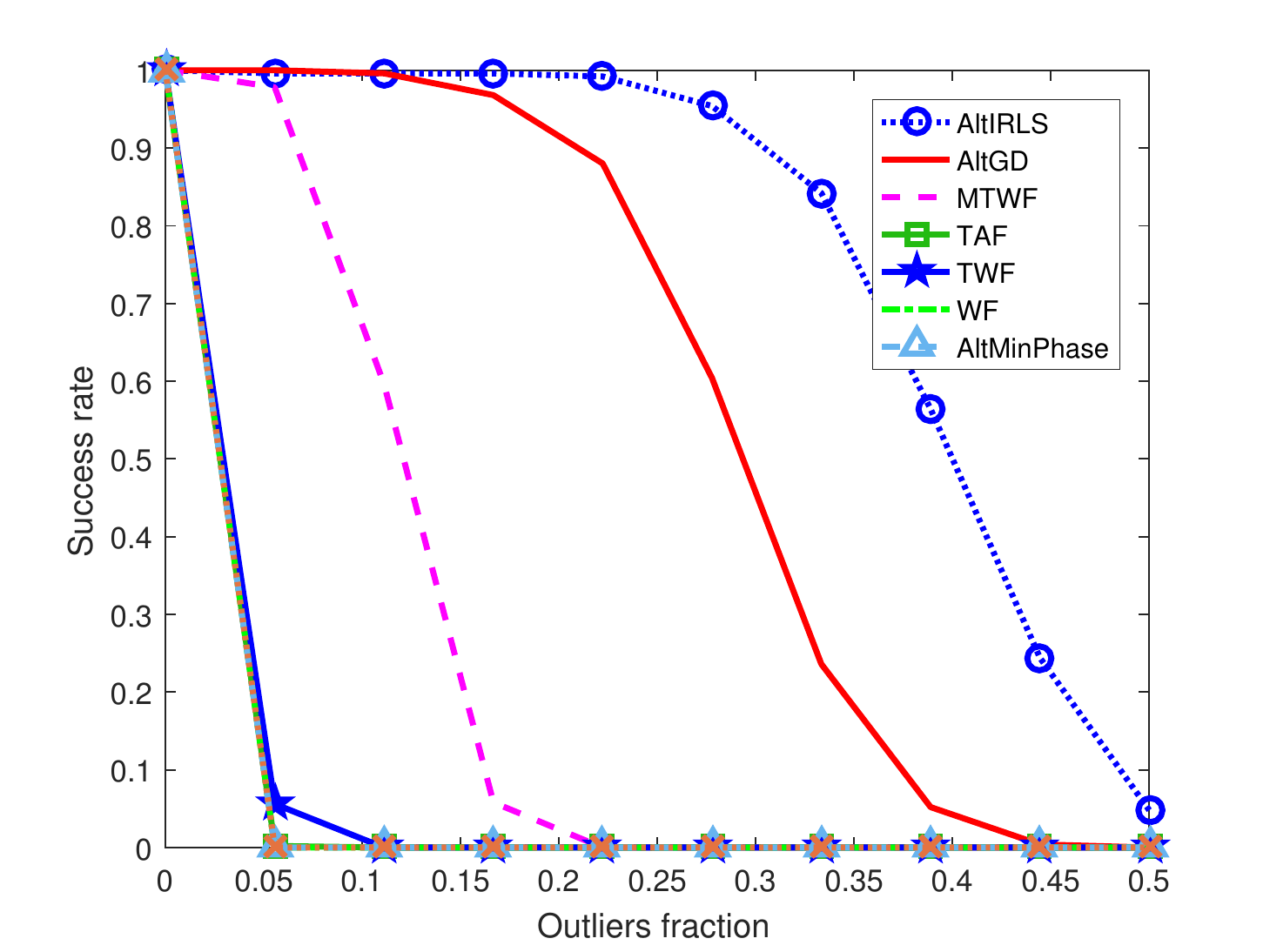}
        \caption{Success rate of exact recovery versus outlier fraction in GMM noise.}\label{Fig_rate}
    \end{center}
\end{figure}

Fig. \ref{Fig_rate} shows the success rate versus the proportion of outliers in the measurements, where the GMM is employed to control the outlier fraction. We vary $c_2$ from 0 to 0.5 and fix $\sigma_1^2=0$ and $\sigma_2^2=100$. SNR is 10 dB. For the proposed methods, we choose $p=0.4$ for comparison.
We declare {\em success} when the squared Euclidean distance of the estimate $\hat\x$ from the ground-truth $\x$ is smaller than or equal to $10^{-4}$; i.e., the success rate is computed as
\begin{align}
\text{success rate} = \frac{1}{500}\sum_{i=1}^{500} \gamma_i
\end{align}
where
\begin{align}
\gamma_i =
\begin{cases}
1,\quad \text{if}\ \|\hat{\x}_i-\x\|_2^2\leq 10^{-4}
\\
0,\quad \text{otherwise.}
\end{cases}
\end{align}
Fig. \ref{Fig_rate} shows that our methods outperform the benchmark methods by a large margin in terms of success rate. When the outlier fraction is greater than 0.25, our methods can still estimate ${\bf x}$ quite accurately, while the other six competitors all fail. Note that the performance of AltIRLS is better than AltGD -- but the complexity of AltGD is much lower.

\begin{figure}
    \begin{center}
        \includegraphics[width=1\linewidth]{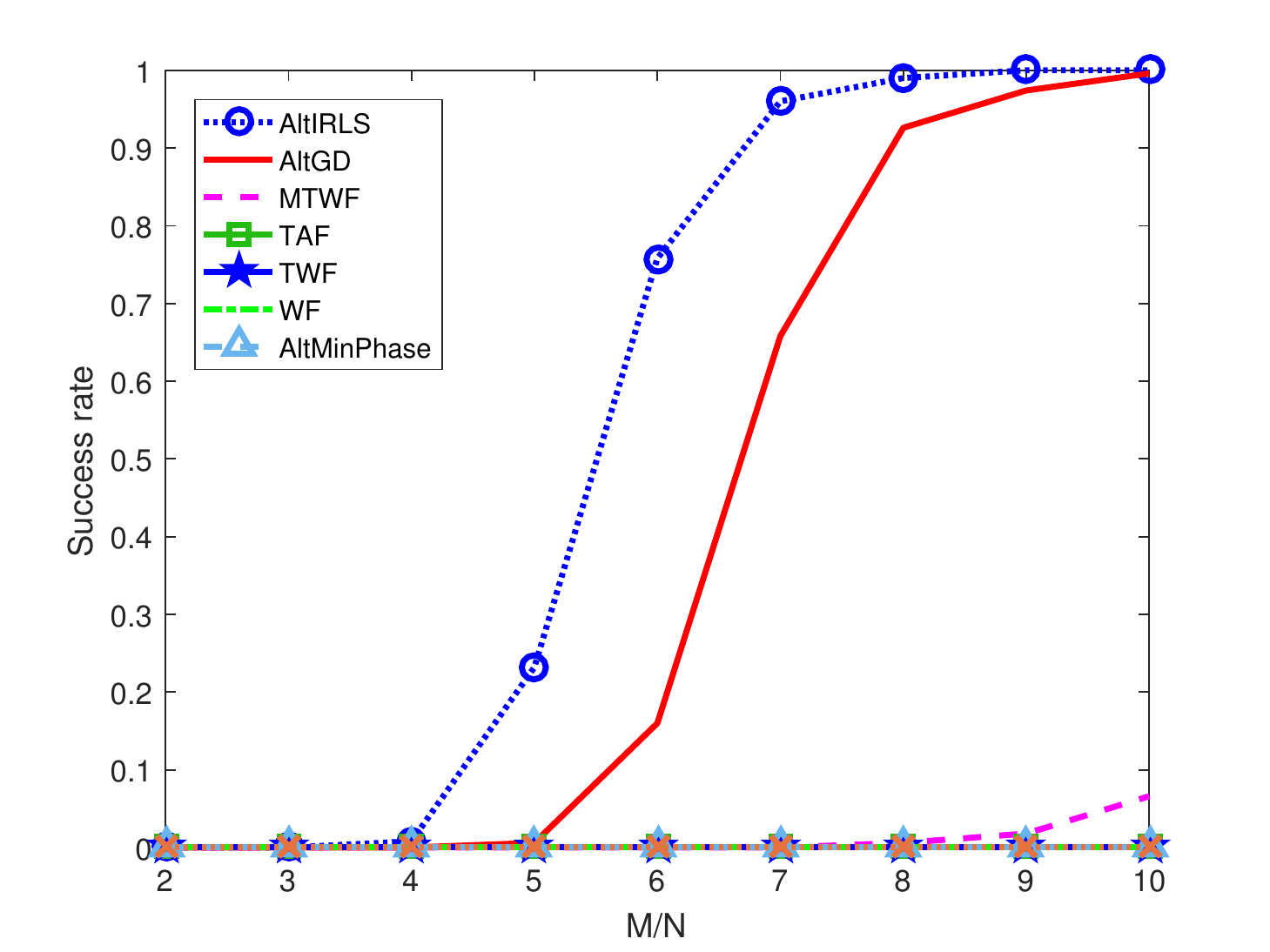}
        \caption{Success rate of exact recovery versus $M/N$ under GMM noise.}\label{Fig_rate2}
    \end{center}
\end{figure}

Fig. \ref{Fig_rate2} plots the success rate versus sample complexity (measured in terms of $M/N$ with $N=16$), where SNR is 10 dB and 20\% of the data are corrupted by outliers. It is obvious that our two algorithms have the highest success rates when $M/N\geq 5$, while the other algorithms have very low success rates even when $M/N\geq 8$.

\subsection{Performance Comparison on Caffeine Molecule Image Data}
\begin{figure*}
    \begin{center}
        \subfigure[Original]{\label{org_mole} \includegraphics[scale=0.45, trim=0 0 0 0]{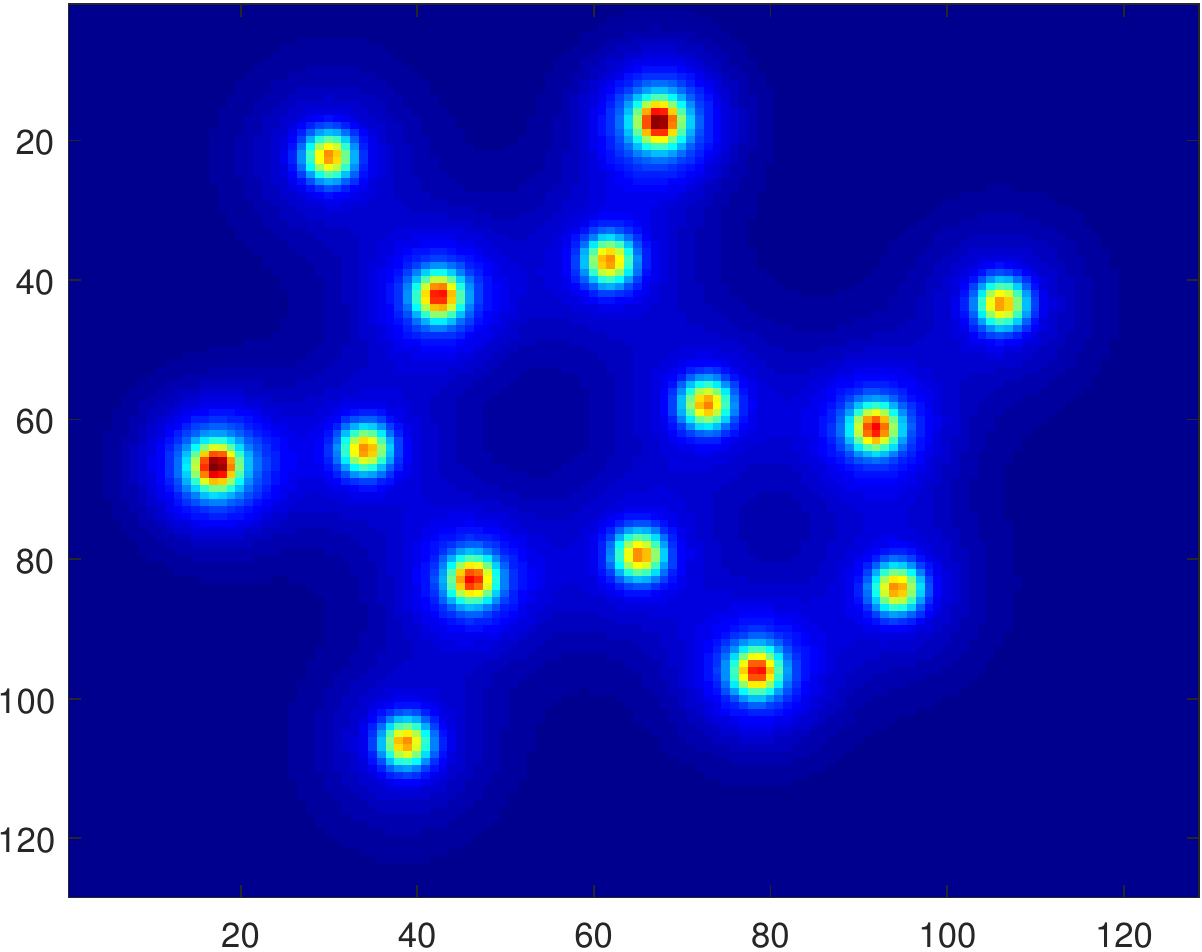}}
        \subfigure[BI-AltGD]{\label{bsg_mole} \includegraphics[scale=0.45,trim=0 0 0 0]{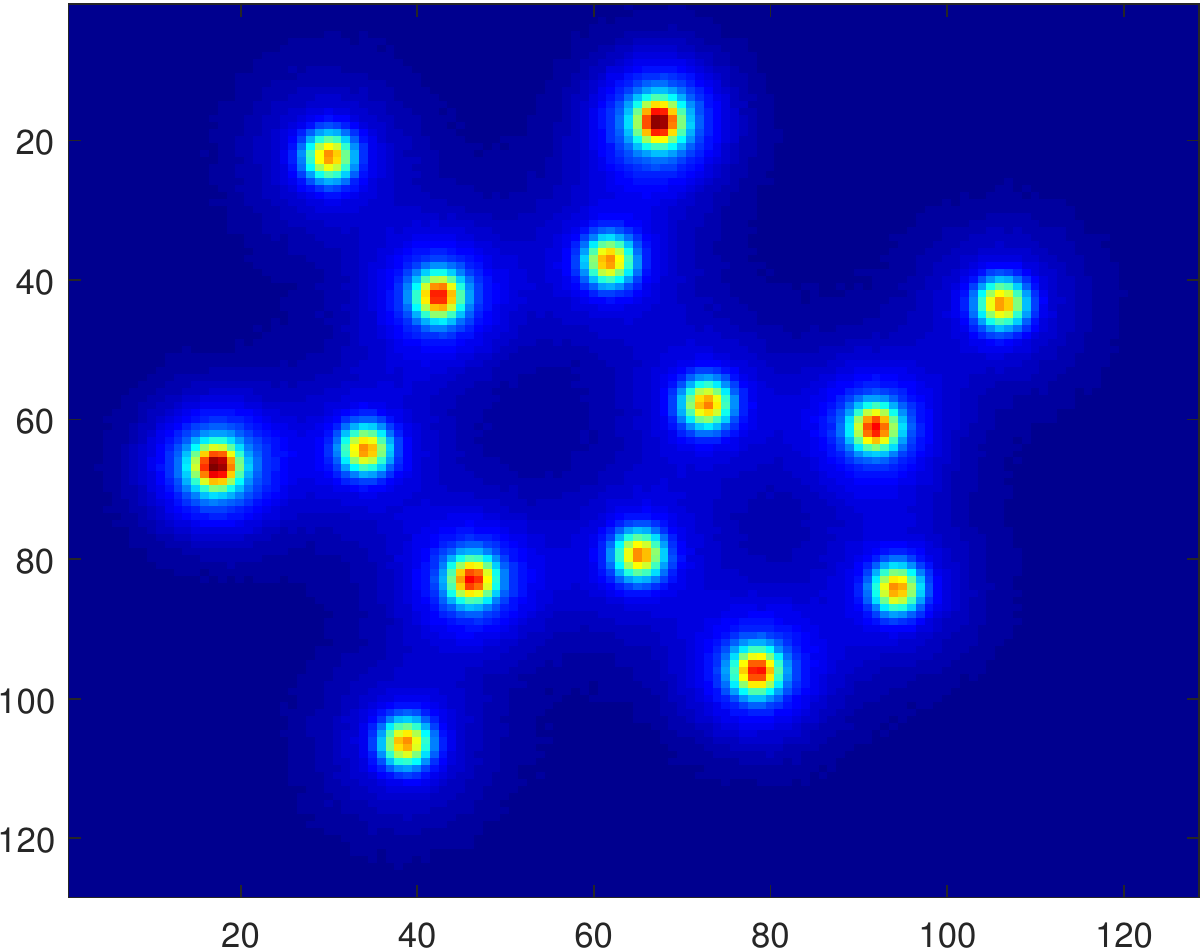}}
        \subfigure[AltGD]{\label{rbsg_mole} \includegraphics[scale=0.45, trim=0 0 0 0]{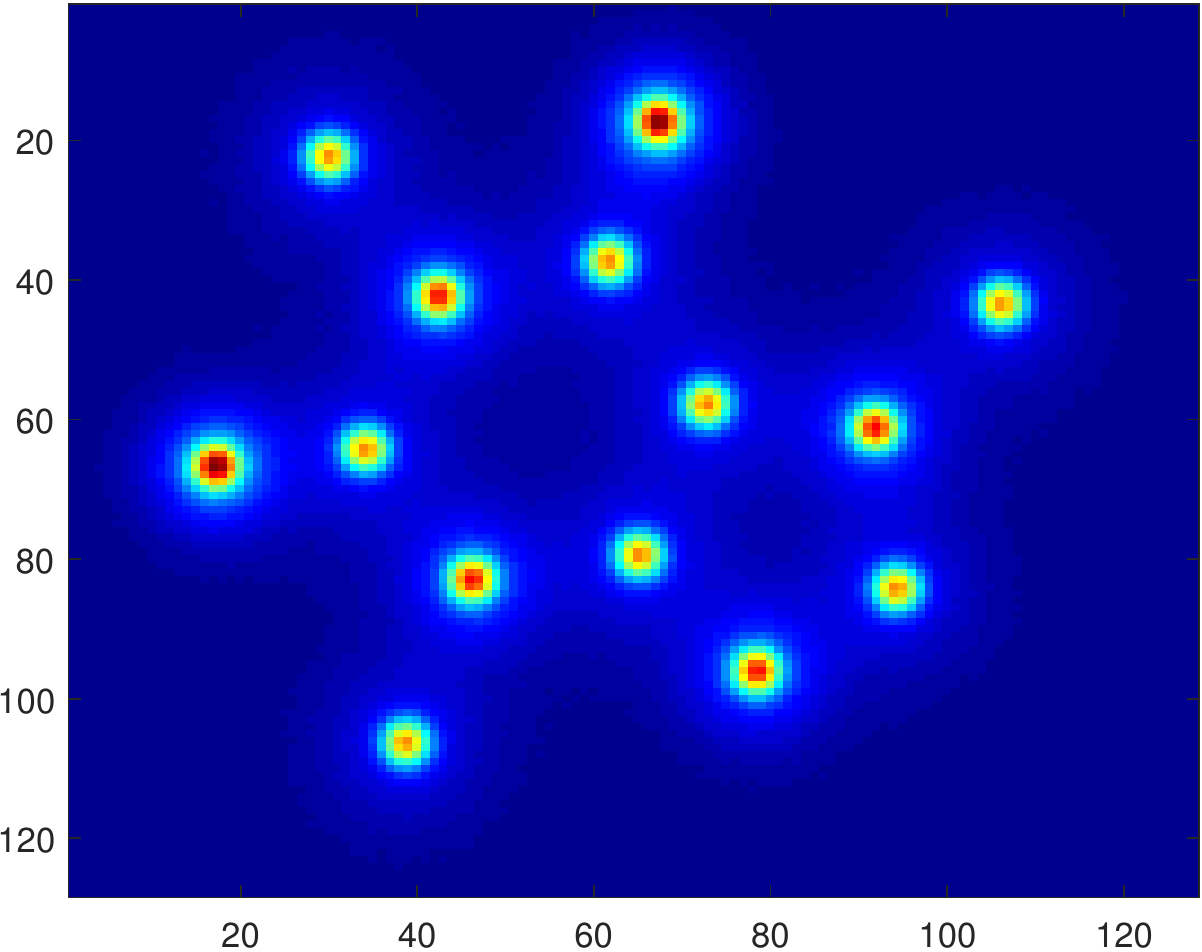}}\\
        \subfigure[TAF]{\label{taf_mole} \includegraphics[scale=0.45,trim=0 0 0 0]{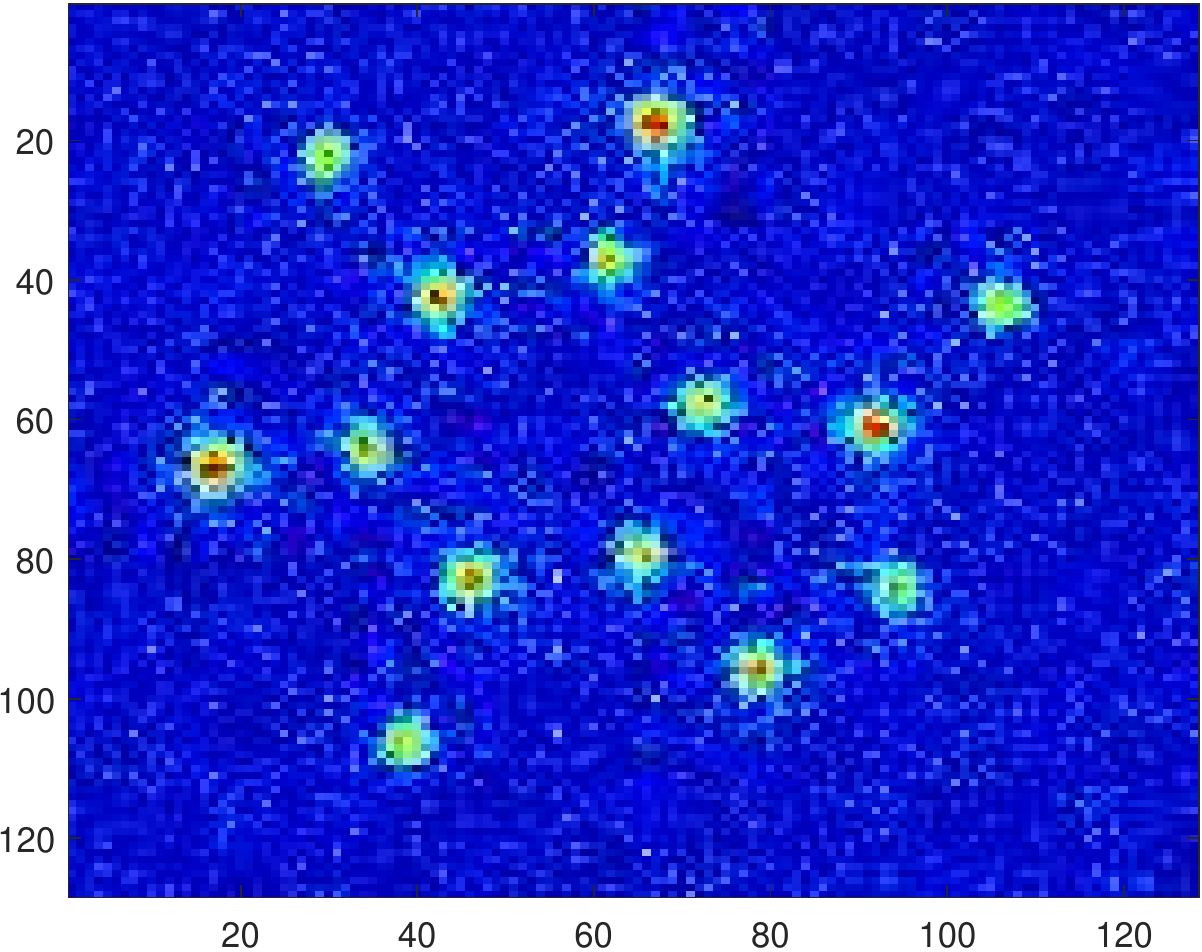}}
        \subfigure[TWF]{\label{twf_mole} \includegraphics[scale=0.45,trim=0 0 0 0]{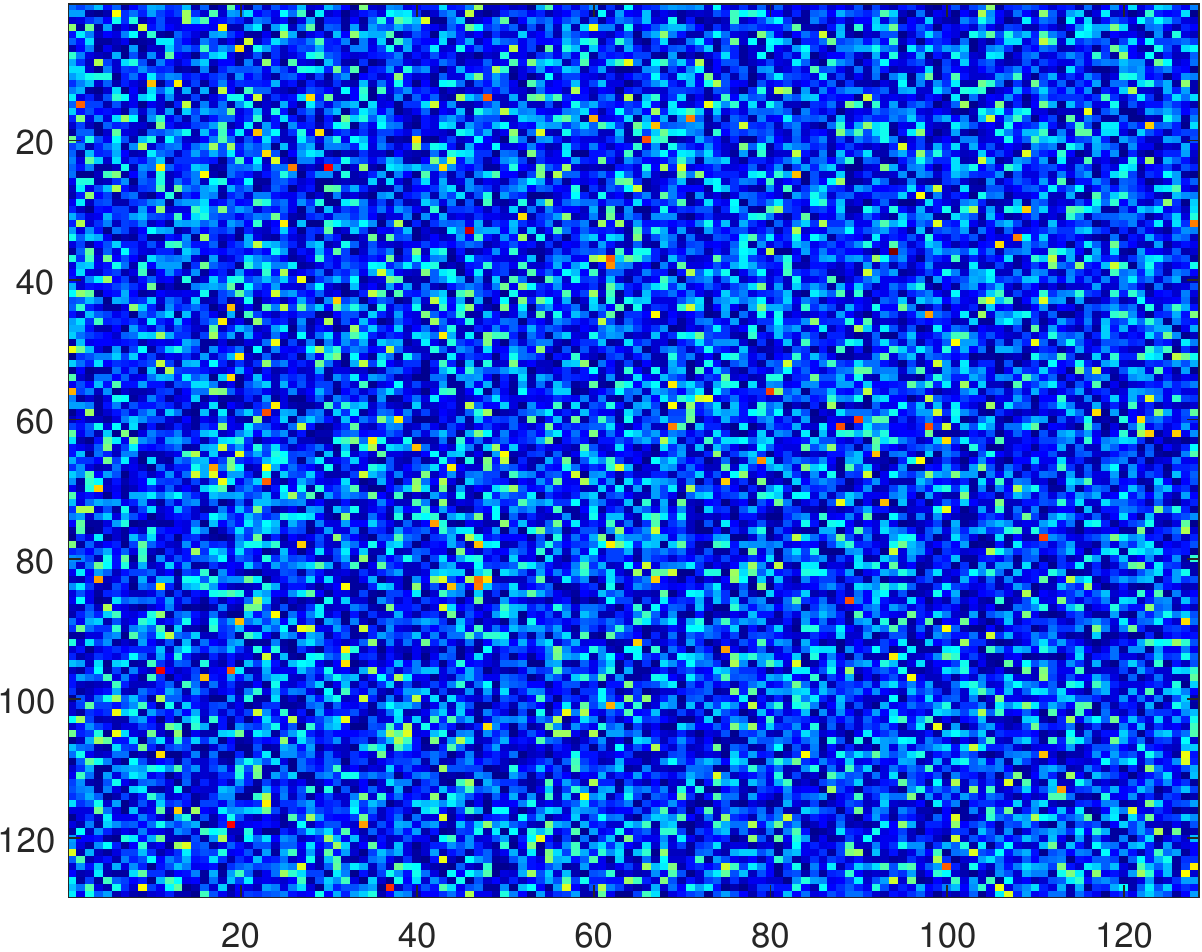}}
        \subfigure[MTWF]{\label{mtwf_mole} \includegraphics[scale=0.45,trim=0 0 0 0]{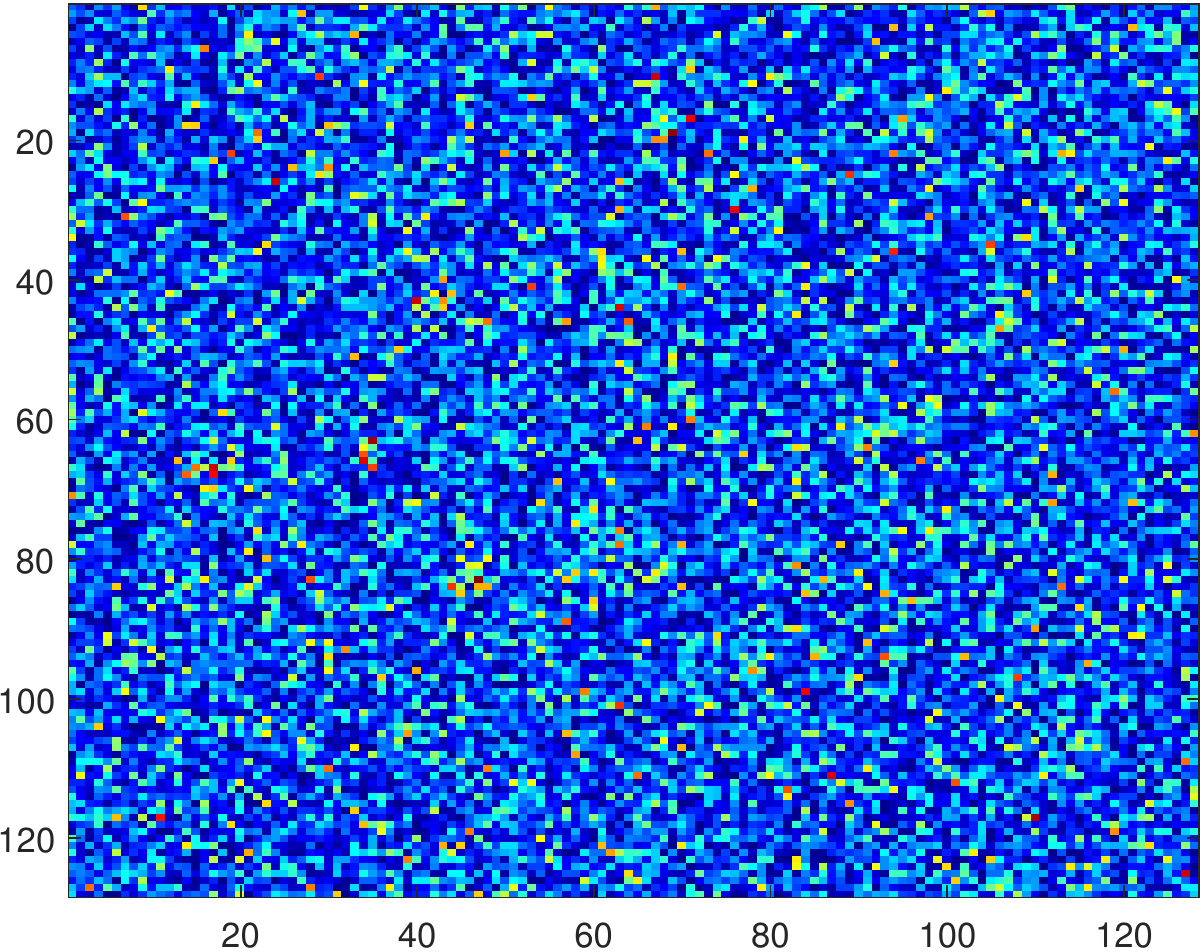}}
        \caption{Retrieving molecule image in S$\alpha$S noise.}
        \label{fig:Molecule}
    \end{center}
\end{figure*}

In this subsection, we showcase the performance using real caffeine molecule data that is often used to test phase retrieval algorithms \cite{WF}. The caffeine molecule data with size $128 \times 128$ is the projection of electron density of a 3-D Caffeine molecule's density map onto the $xy$-plane. The objective is to reconstruct the data from the magnitude of its masked Fourier transform. Here, we use the same measuring process as in the previous simulations and
$K = 8$ masks are used to generate measurements. We compare the performance of AltGD, TAF, MTWF, and TWF. 
We also include the block incremental (BI) implementation of AltGD (cf. Section II.C2), which is referred to as \emph{BI-AltGD}.
The outliers are generated from GMM with $c_2=0.3$, $\sigma_1^2=0$ and $\sigma_2^2=100$, which indicates that there are $30\%$ of the data are corrupted by outliers. SNR is 0 dB and the data is normalized with unit norm.
Fig. \ref{fig:Molecule} plots the retrieved molecule's density map using AltGD, BI-AltGD with $p=1.3$, TAF, MTWF and TWF. It is seen in Fig. \ref{fig:Molecule} that our schemes still work well, while the other competitors yield blurred images.

\subsection{Performance Comparison with 2D Fourier Measurements}\label{sec:2dfourier}
In the above simulations, the measurement vectors are generated from coded diffraction pattern (CDP) with random masks. The success of the existing algorithms such as the well-known PhaseLift, WF and its variants are all based on the randomness of measurements. However, to the best of our knowledge, the CDP model is still considered to be impractical, as there has been no device supporting the masking technique yet. In practice, most devices record the intensity of the Fourier transform of the object, where the intensity is usually proportional to the magnitude of the oversampled 2D Fourier measurements with $2 \times$ oversampling. In this case, the measurement matrix $\A$ is a Kronecker product of two oversampled Fourier matrices.

With 2D Fourier measurements, the most successful method is Fienup's hybrid-input-output (HIO) algorithm \cite{Fienup}, which has been observed to give empirically reliable phase retrieval performance, especially in the noiseless case.
In the noisy case, HIO can also provide good initialization for other algorithms, e.g., GS. Therefore, HIO and HIO-initialized GS are employed as baselines in this subsection.

\begin{figure}[h!]
    \begin{center}
        \includegraphics[width=1\linewidth]{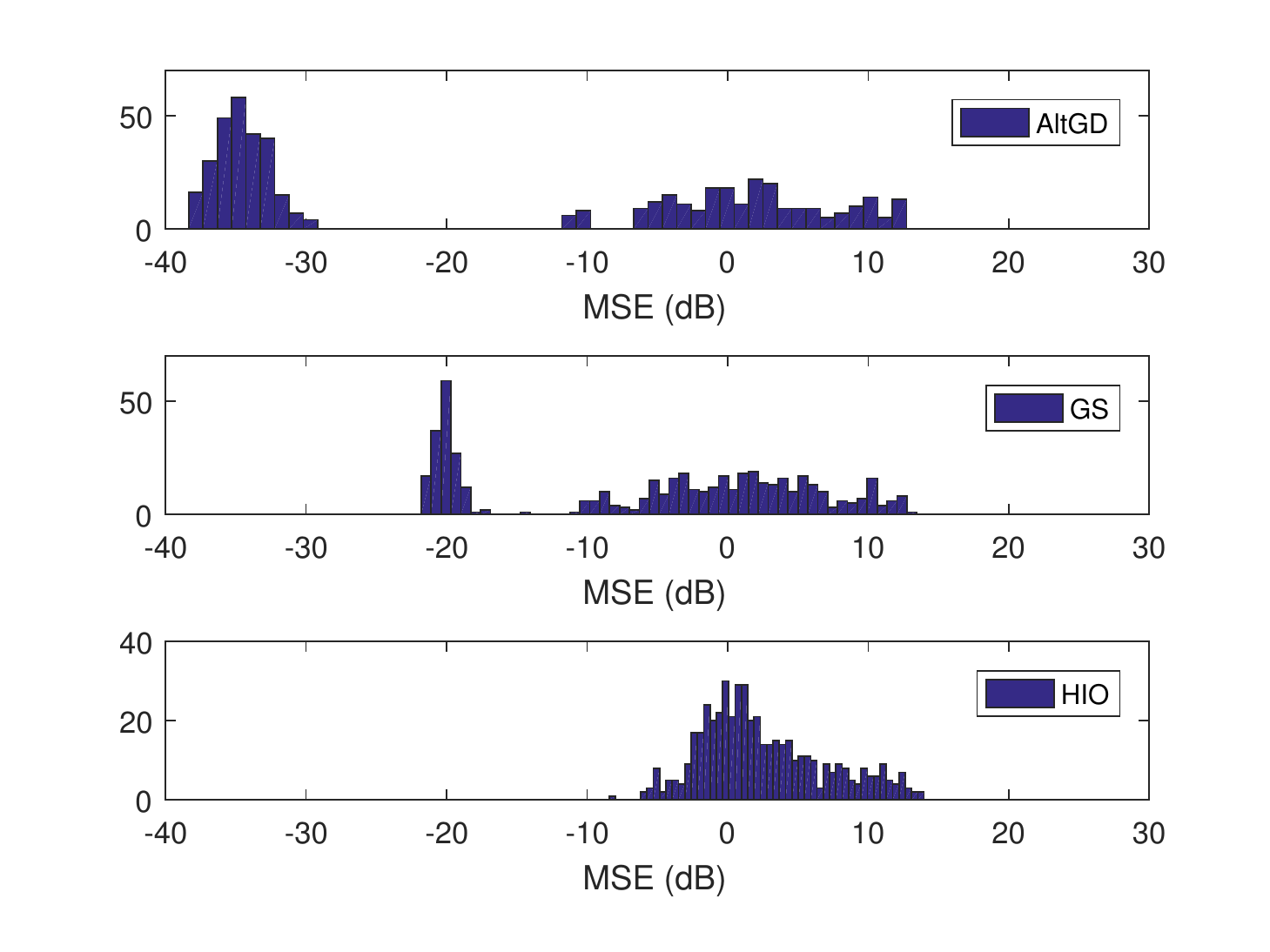}
        \caption{Signal recovery performance comparison with 2D Fourier measurements in GMM noise.}\label{Fig_mse_2dFFT}
    \end{center}
\end{figure}

In this simulation, we compare the performance of AltGD, HIO and GS under 2D Fourier measurements in the presence of outliers, where GMM-type outliers are generated with $c_2=0.1$, $\sigma_1^2=0$ and $\sigma_2^2=100$. Each entry of the signal matrix $\X\in\bR^{16\times16}$ is generated from the normal distribution. After zero-padding around $\X$, we have $\tilde{\X}\in\bR^{32\times32}$. Then the measurements are obtained by first calculating 2D fast Fourier transformation (FFT) on $\tilde{\X}$ and then  taking the magnitude of $\mathrm{2D\, FFT}(\tilde{\X})$, where the sizes of left and right Fourier matrices are both $32\times32$. Note that SNR is 10 dB and is computed as $10\log_{10}(\|\tilde\X\|_F^2/\|\N\|_F^2)$ where $\N\in\bR^{32\times32}$ is the noise matrix.  For AltGD and GS, HIO with 5000 iterations is employed to provide an initial estimate of $\X$, and then an additional 5000 iterations are used in AltGD and GS, respectively. Moreover, HIO with 10000 iterations is included as a baseline. For AltGD, we choose $p=1.3$. The MSEs obtained from 1000 Monte-Carlo tests are plotted in Fig. \ref{Fig_mse_2dFFT}. It is seen that HIO does not perform very well, and is inferior to AltGD and GS. AltGD has better performance than GS, since most of its MSEs are concentrated around $-35$ dB, while those obtained by GS are located around $-20$ dB.

\begin{figure}
    \begin{center}
        \subfigure[Original]{\label{org_cameraman} \includegraphics[scale=0.9, trim=0 0 0 0]{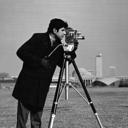}}
        \subfigure[AltGD]{\label{cameraman_altgd} \includegraphics[scale=0.9,trim=0 0 0 0]{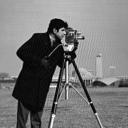}}\\
        \subfigure[GS]{\label{cameraman_gs} \includegraphics[scale=0.9, trim=0 0 0 0]{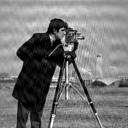}}
        \subfigure[HIO]{\label{cameraman_hio} \includegraphics[scale=0.9,trim=0 0 0 0]{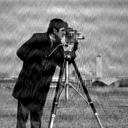}}
        \caption{Retrieving Cameraman image in GMM noise with 2D Fourier measurements.}
        \label{fig:cameraman}
    \end{center}
\end{figure}

Finally, Fig. \ref{fig:cameraman} shows the recovered image of $128\times128$ Cameraman from 2D Fourier measurements under GMM noise, where SNR is 12 dB and the noise is generated in the same way as for Fig. \ref{Fig_mse_2dFFT}. The number of iterations used for initialization and refinement are all 5000. For AltGD, we set $p=0.6$. It is seen that the image recovered from AltGD is much more clear than those recovered by HIO and GS.


\section{Conclusion}

In this paper, we considered phase retrieval in the presence of grossly corrupted data -- i.e., outliers.
We formulated this problem as an $\ell_p$ fitting problem, where $0<\ell_p<2$,
and provided an algorithmic framework that is based on two-block inexact alternating optimization.
Two algorithms, namely, AltIRLS and AltGD, were proposed under this framework.
Although the algorithms cannot be analyzed using standard convergence results for alternating optimization due to a nonconvex constraint, we managed to show that the algorithms converge to a KKT point. The tools used for convergence analysis can also be used for analyzing convergence of other types of algorithms that involve non-convex constraints and inexact alternating optimization.
Pertinent CRBs were derived for the noisy measurement models considered.
Simulations showed that the proposed algorithms are promising in dealing with outliers in the context of phase retrieval using both random measurements and 2D Fourier measurements, for simulated and real image data.

\appendices
\section{Proof of Proposition~\ref{prop:convergence}}\label{Appendix_convergence}
Let us denote
\begin{equation*}
{f({\bf x},{\bf u})} = \sum_{m=1}^{M}\left(|y_mu_m - \a_m^H\x |^2 + \epsilon\right)^{p/2}.
\end{equation*}
To simplify the analysis, let us define
\begin{equation*}
\begin{aligned}
\tilde{\bf u}_m&=[{\rm Re}\{u_m\},{\rm Im}\{u_m\}]^T,\\
\tilde{\bf x}&=[{\rm Re}\{{\bf x}^T\},{\rm Im}\{{\bf x}^T\}]^T,\\
\tilde{\bf A}_m&=\begin{bmatrix}
{\rm Re}\{{\bm a}_m^H\}, &- {\rm Im}\{{\bm a}_m^H\}\\
{\rm Im}\{{\bm a}_m^H\}, & {\rm Re}\{{\bm a}_m^H\}
\end{bmatrix}.
\end{aligned}
\end{equation*}
Therefore, our problem can be written as
\begin{equation}\label{eq:newproblem}
\begin{aligned}
\min_{\{\tilde{\bf u}_m\},\tilde{\bf x}}~& { f(\tilde{\bf x},\{\tilde{\bf u}_m\}) } = \sum_{m=1}^{M}\left(\|y_m \tilde{\bf u}_m - \tilde{\bf A}_m\tilde{\x} \|_2^2 + \epsilon\right)^{p/2}\\
{\rm s.t.}~& \|\tilde{\bf u}_m\|_2^2 = 1,~\forall m.
\end{aligned}
\end{equation}
Note that in the new expression \eqref{eq:newproblem}, all the variables are real-valued.
Accordingly, we may define
\begin{equation}
{ g(\tilde{\bf x}, \{\tilde{\bf u}_m^{(r)}\})} = \sum_{m=1}^{M}\left(w_m^{(r)}\|y_m \tilde{\bf u}_m^{(r)} - \tilde{\bf A}_m\tilde{\x} \|_2^2 + \phi(w_m^{(r)})\right).
\end{equation}
Now our algorithm can be re-expressed as
\begin{align}
\tilde{\bf x}^{(r+1)} &= \arg\min_{\tilde{\bm x}}~ { g(\tilde{\bf x}, \{\tilde{\bf u}^{(r)}_m\}) } \label{eq:updaterule1} \\
\tilde{\bf u}_m^{(r+1)}& = \arg\min_{\|\tilde{\bf u}_m\|_2^2=1,~\forall m}~ { f(\tilde{\bf x}^{(r+1)}, \{\tilde{\bf u}_m\}) } \label{eq:updaterule2}
\end{align}

Let us assume the boundedness of $\tilde{\x}^{(r)}$ and $\tilde{\bf u}^{(r)}$ and $w_m^{(r)}$, which will be shown later. Under this assumption,
the gradients of $ g(\tilde{\bf x}, \{\tilde{\bf u}_m^{(r)}\})$ and ${ f(\tilde{\bf x},\{\tilde{\bf u}_m\}) } $ exist.
According to Lemma~\ref{lem:conjugate}, it is easily seen that
{
\begin{subequations}\label{eq:g_grad}
	\begin{align}
	f(\tilde{\bf x},\{\tilde{\bf u}_m^{(r)}\}) &\leq 	g(\tilde{\bf x},\{\tilde{\bf u}_m^{(r)}\})\quad \forall \tilde{\bf x}, \label{eq:property1} \\
	f(\tilde{\bf x}^{(r)},\{\tilde{\bf u}_m^{(r)}\}) &= 	g(\tilde{\bf x}^{(r)},\{\tilde{\bf u}_m^{(r)}\}) \label{eq:property2}\\
	\nabla_{\tilde{\bf x}} f(\tilde{\bf x}^{(r)},\{\tilde{\bf u}_m^{(r)}\}) & = \nabla_{\tilde{\bf x}} g(\tilde{\bf x}^{(r)},\{\tilde{\bf u}_m^{(r)}\}).  \label{eq:property3}
	\end{align}
\end{subequations}
}

Now, we have
{
\begin{subequations}
	\begin{align}
	f(\tilde{\bf x}^{(r)},\{\tilde{\bf u}_m^{(r)}\})& = g(\tilde{\bf x}^{(r)},\{\tilde{\bf u}_m^{(r)}\}) \label{eq:st1}\\
	& \geq g(\tilde{\bf x}^{(r+1)},\{\tilde{\bf u}_m^{(r)}\})  \label{eq:st2}\\
	&\geq f(\tilde{\bf x}^{(r+1)},\{\tilde{\bf u}_m^{(r)}\})  \label{eq:st3} \\
	&\geq f(\tilde{\bf x}^{(r+1)},\{\tilde{\bf u}_m^{(r+1)}\})  \label{eq:st4}
	\end{align}
\end{subequations}
}
where \eqref{eq:st1} follows \eqref{eq:property2},
\eqref{eq:st2} is obtained because of \eqref{eq:updaterule1},
\eqref{eq:st3} holds since we have the property in \eqref{eq:property1},
and \eqref{eq:st2} is obtained by the fact that the subproblem w.r.t. ${\bf u}$ is optimally solved, i.e., \eqref{eq:updaterule2}.

Assume that $\{r_j\}$ denotes the index set of a convergent subsequence,
and that $\{\tilde{\bf x}^{(r_j)},\{\tilde{\bf u}_m^{(r_j)}\}\}$ converges to $(\tilde{\bf x}^\star,\{\tilde{\bf u}_m^\star\})$ Then, we have
\begin{subequations}
	\begin{align}
	g(\tilde{\bf x},\{\tilde{\bf u}_m^{(r_j)}\}) &\geq g(\tilde{\bf x}^{(r_j+1)},\{\tilde{\bf u}_m^{(r_j)}\}) \label{eq:st11}\\
	&\geq f(\tilde{\bf x}^{(r_j+1)},\{\tilde{\bf u}_m^{(r_j)}\}) \label{eq:st22}\\
	&\geq f(\tilde{\bf x}^{(r_j+1)},\{\tilde{\bf u}_m^{(r_j+1)}\}) \label{eq:st33}\\
	&\geq f(\tilde{\bf x}^{(r_{j+1})},\{\tilde{\bf u}_m^{(r_{j+1})}\}) \label{eq:st44}\\
	& = g(\tilde{\bf x}^{(r_{j+1})},\{\tilde{\bf u}_m^{(r_{j+1})}\}), \label{eq:st55}
	\end{align}
\end{subequations}
where \eqref{eq:st44} is obtained by the fact that $r_{j+1}\geq r_j+1$ since $r_j$ indexes a subsequence.
Taking $j\rightarrow\infty$, and by the boundedness of ${\bf W}^{(r)}$ and continuity of $g(\cdot)$, we see that
\begin{equation}\label{eq:gmin}
g(\tilde{\bf x},\{\tilde{\bf u}_m^\star\})\geq g(\tilde{\bf x}^\star,{\bf u}^{\star}).
\end{equation}
The inequality in \eqref{eq:gmin} means that $\tilde{\bf x}^\star$ is blockwise minimizer of $g(\tilde{\bf x},\{\tilde{\bf u}_m^\star\})$.
Therefore, it satisfies the partial KKT condition w.r.t. $\tilde{\bf x}$, i.e.,
\begin{equation}
\nabla_{\tilde{\bf x}} g(\tilde{\bf x}^\star,\{\tilde{\bf u}_m^\star\}) = {\bf 0}.
\end{equation}
By \eqref{eq:property3}, we immediately have
\begin{equation}\label{eq:KKT1}
\nabla_{\tilde{\bf x}} f(\tilde{\bf x}^\star,\{\tilde{\bf u}_m^\star\}) = {\bf 0}.
\end{equation}

Similarly, by the update rule in \eqref{eq:updaterule2}, we have
\begin{align}
f(\tilde{\bf x}^{(r_j)},{\bf u}) \geq f(\tilde{\bf x}^{(r_j)},\{\tilde{\bf u}_m^{(r_j)}\})
\end{align}
and thus
\begin{equation}
f(\tilde{\bf x}^{\star},\{{\bf u}_m\})\geq f(\tilde{\bf x}^{{\star}},\{{\bf u}_m^{{\star}}\}).
\end{equation}
Therefore, $\tilde{\bf u}_m^\star$ also satisfies the partial KKT condition
\begin{equation}\label{eq:KKT2}
\nabla_{\tilde{\bf u}_m}~f(\tilde{\bf x}^\star,\{\tilde{\bf u}_m^\star\}) + {\bm \lambda}_m^\star \odot \tilde{\bf u}_m^\star={\bm 0},~\forall m
\end{equation}
where ${\bm \lambda}_m\in\mathbb{R}^2$ are dual variables.
{Combining \eqref{eq:KKT1} and \eqref{eq:KKT2}, it follows that every limit point of the solution sequence is a KKT point.}

Now, we rigorously show the boundedness of $\tilde{\x}^{(r)}$ and $\tilde{\bf u}^{(r)}$ and $\W^{(r)}$. 
Let us first show that $\W^{(r)}$ is bounded given finite $\tilde{\x}^{(r)}$ and $\tilde{\bf u}^{(r)}$.
This is relatively easy, since ${\bf y}$ and ${\bf A}$ are always bounded and thus ${\bf W}^{(r)}$ is also bounded when $0<p<2$ and $\epsilon>0$ by definition (cf. Eq.~\eqref{eq:w_def}).
Now, we show that $\tilde{\x}^{(r)}$ and $\tilde{\bf u}^{(r)}$ are bounded.
Since the algorithm is essentially majorization minimization, there is no risk for $\tilde{\x}^{(r)}$ being unbounded, if the initialization is bounded, $\A$ has full column rank, and $\y$ is bounded.
To explain, one can see that
	\begin{equation}
	\begin{aligned}
 \|\y\odot\tilde{\bf u} ^{(r)}- \tilde{\A}\tilde{ \x}^{(0)}\|_p^p & \geq \|\y\odot\tilde{\bf u} ^{(r)} - \tilde{\A}\tilde{ \x}^{(r)}\|_p^p \\
	& \geq | \|\y\|_p^p - \| \tilde{\A}\tilde{ \x}^{(r)}\|_p^p|\\
	& \geq \|\tilde{\A}\tilde{ \x}^{(r)}\|_p^p - \|\y\|_p^p.
	\end{aligned}
	\end{equation}
	The first inequality holds because block upper bound minimization algorithm always decreases the objective \cite{bsum}.
	The second inequality is due to the triangle inequality for $\|\cdot\|_p^p$ when $0<p<1$ \cite{boundedness}; for $p>1$, one can apply the same argument to $\|\cdot\|_p$.
	From the above, one can see that
	\begin{equation*}
	\| \tilde{\A}\tilde{ \x}^{(r)}\|_p^p  \leq   \infty . 
\end{equation*}   
Therefore, if $\A$ does not have a null space, then $\tilde{\x}^{(r)}$ has to be bounded. Consequently, $\W^{(r)}$ is always bounded from above.

The next step is to show that the whole sequence converges to ${\cal K}$, the set of all the KKT points. This is relatively straightforward. Let us assume that the whole sequence does not converge to ${\cal K}$. This means that there exists a subsequence which does not converge to ${\cal K}$.
The boundedness of $\tilde{\x}^{(r)},\tilde{\bf u}^{(r)}$ implies that every subsequence of $\{\tilde{\x}^{(r)},\tilde{\bf u}^{(r)}\}$ has a limit point. We have just shown that every limit point is a KKT point, and thus this is a contradiction. Therefore, the whole sequence has to converge to ${\cal K}$.


\section{CRB for Laplacian Noise}\label{Appendix_CRB_Laplacian}
The likelihood function for Laplacian noise is given by \cite{LapNoiseCRB}-\cite{LapNoiseModel}
\begin{align}\label{pdf_lap}
  p(\y;\x) = \prod_{i=1}^M\frac{1}{\sqrt{2}\sigma_n}\exp\left\{ -\frac{\sqrt{2}}{\sigma_n}\left|y_i - |\a_i^H\x|\right| \right\}
\end{align}
where the noise variance is $\sigma_n^2$, and its log-likelihood function is
\begin{align}
    \ln p(\y;\x) = -M\ln(\sqrt{2}\sigma_n) - \frac{\sqrt{2}}{\sigma_n}\sum_{i=1}^M\left|y_i-|\a_i^H\x|\right|.
\end{align}
The vector of unknown parameters for complex-valued $\x$ is
\begin{align}\label{A2:beta}
  \bbe = [\Re\{x_1\}\ \cdots\ \Re\{x_N\},\ \Im\{x_1\}\ \cdots\ \Im\{x_N\}]^T.
\end{align}
It is worth noting that $\sigma_n^2$ is actually an unknown parameter which should be considered as a part of $\bbe$. However, since $\sigma_n^2$ is uncorrelated with the real and imaginary parts of $x_i$, their mutual Fisher information is zero. It will not impact the final CRB formula for $\x$. For this reason, we do not include $\sigma_n^2$ in $\bbe$.
Thus, the FIM can be partitioned into four parts, i.e.,
\begin{align}\label{CRB:F}
    \F_{L,c} =
    \begin{bmatrix}
    \F_{L,rr} & \F_{L,ri} \\
    \F_{L,ir} & \F_{L,ii}
    \end{bmatrix}
\end{align}
where
\begin{align} \label{LAP:FIM}
  [\F_{L,c}]_{m,n} = \mathbb{E}\left[ \frac{\partial \ln p(\y;\x)}{\partial\bbe_m} \frac{\partial \ln p(\y;\x)}{\partial\bbe_n}  \right].
\end{align}
The partial derivative of $\ln p(\y;\x)$ with respective to $\bbe_m$ is
\begin{align}
  \frac{\partial \ln p(\y;\x)}{\partial\bbe_m} =&\, -\frac{\sqrt{2}}{\sigma_n}\sum_{i=1}^M\frac{\partial \left|y_i - |\a_i^H\x|\right|}{\partial \bbe_m} \notag\\
    =&\, \frac{\sqrt{2}}{\sigma_n}\sum_{i=1}^M \frac{y_i - |\a_i^H\x|}{\left|y_i - |\a_i^H\x|\right|} \frac{\partial |\a_i^H\x|}{\partial \bbe_m} \notag\\
    =&\, \frac{\sqrt{2}}{\sigma_n}\sum_{i=1}^M \sgn(y_i - |\a_i^H\x|) \frac{\partial |\a_i^H\x|}{\partial \bbe_m} \label{LAP:pxrm}
\end{align}
where
\begin{align}
  \sgn(a) = \left\{ \begin{aligned} 1&, \quad a>0 \\ -1&,\quad a<0 \end{aligned} \right.
\end{align}
and
\begin{equation}\label{CRB:4}
\frac{\partial |\a_i^H\x|}{\partial \bbe_m} = \left\{
\begin{aligned}
&\frac{[\Re\left\{\a_i\a_i^H\x\right\}]_m}{|\a_i^H\x|},\ \text{for}\ \bbe_m = \Re\{x_m\}\\
&\frac{[\Im\left\{\a_i\a_i^H\x\right\}]_m}{|\a_i^H\x|} ,\ \text{for}\ \bbe_m = \Im\{x_m\}.
\end{aligned}
\right.
\end{equation}

Substituting \eqref{LAP:pxrm} into \eqref{LAP:FIM}, we have
\begin{align}
  [\F_{L,c}]_{m,n} =&\, \frac{2}{\sigma_n^2} \sum_{i=1}^M\sum_{j=1}^M \frac{\partial |\a_i^H\x|}{\partial \bbe_m}\frac{\partial |\a_i^H\x|}{\partial \bbe_n} \notag\\
    & \times \mathbb{E}\left[\sgn(y_i - |\a_i^H\x|)\cdot\sgn(y_j - |\a_j^H\x|)\right]. \label{LAP:F111}
\end{align}
Next we compute the value of $\mathbb{E}[\sgn(y_i - |\a_i^H\x|)\sgn(y_j - |\a_j^H\x|)]$. For notational simplicity, let $s_i = \sgn(y_i - |\a_i^H\x|)$. It is obvious that when $i=j$, we have
\begin{align}
  \mathbb{E}\left[s_is_j\right] = 1. \label{sisj_ieqj}
\end{align}
For $i\neq j$, we first write
\begin{align}
  \mathbb{E}\left[s_is_j\right] =&\ \pr(s_i=s_j)\times(+1) + \pr(s_i\neq s_j)\times(-1)\notag\\
    =&\ 2\pr(s_i=s_j) - 1 \label{sisj_ineqj}
\end{align}
where $\pr$ standards for the probability.
Then the value of $\pr(s_i=s_j)$ is computed as
\begin{align}
  \pr(s_i=s_j) =&\ \pr(s_i=1|s_j=1)\pr(s_j=1) \notag\\
   & + \pr(s_i=-1|s_j=-1)\pr(s_j=-1) \notag\\
   =&\ \pr(s_i=1)\pr(s_j=1) \notag\\
   & + \pr(s_i=-1)\pr(s_j=-1)\notag\\
   =&\ 0.5\label{prob}
\end{align}
where $\pr(s_i=1) = \pr(s_i=-1) = 0.5$ and the second equation follows by independence of $s_i$ and $s_j$ when $i\neq j$. Substituting \eqref{prob} into \eqref{sisj_ineqj} and using \eqref{sisj_ieqj} yields
\begin{align}
  \mathbb{E}\left[s_is_j\right] = \left\{ \begin{aligned} 1,&\quad i=j \\ 0,&\quad i\neq j. \end{aligned} \right. \label{sisj_final}
\end{align}
Substituting \eqref{sisj_final} into \eqref{LAP:F111}, we obtain
\begin{align}
  [\F_{L,c}]_{m,n} =&\ \frac{2}{\sigma_n^2} \sum_{i=1}^M \frac{\partial |\a_i^H\x|}{\partial \bbe_m}\frac{\partial |\a_i^H\x|}{\partial \bbe_n}. \label{LAP:F}
\end{align}

Now using \eqref{CRB:4} and \eqref{LAP:F}, the four sub-FIMs can be easily derived as
\begin{align}
  \F_{L,rr} =&\ \frac{2}{\sigma_n^2}\Re\{\A^H\diag(\A\x)\}\cdot\diag(|\A\x|^{-2}) \notag\\
          &\qquad \times\Re\{\A^H\diag(\A\x)\}^T\label{LAP:F11} \\
  \F_{L,ii} =&\ \frac{2}{\sigma_n^2}\Im\{\A^H\diag(\A\x)\}\cdot\diag(|\A\x|^{-2}) \notag\\
          &\qquad \times\Im\{\A^H\diag(\A\x)\}^T\label{LAP:F22} \\
  \F_{L,ri} =&\ \frac{2}{\sigma_n^2}\Re\{\A^H\diag(\A\x)\}\cdot\diag(|\A\x|^{-2}) \notag\\
          &\qquad \times\Im\{\A^H\diag(\A\x)\}^T\label{LAP:F12} \\
  \F_{L,ir} =&\ \F_{L,ri}^T \label{LAP:F21}.
\end{align}
Combining \eqref{LAP:F11}-\eqref{LAP:F21}, we obtain
\begin{align}
  \F_{L,c} = \frac{2}{\sigma_n^2}\G_{L,c}\,\diag(|\A\x|^{-2})\,\G^T_{L,c}
\end{align}
where
\begin{align}
	\G_{L,c} =
	\begin{bmatrix}
	  \Re\{\A^H\diag(\A\x)\}\\
	  \Im\{\A^H\diag(\A\x)\}
	\end{bmatrix}.
\end{align}
This completes the proof of Proposition \ref{proposition1}.

\section{Proof of Rank Property of $\F_{L,c}$ and $\F_{L,r}$}\label{Appendix_F_rank}
We note here that $\F_{L,c}$ is derived under the assumption of nonzero ${\bf a}_m^H\x$. Therefore, $\diag(|\A\x|^{-2})$ is full rank. As a result, computing the rank of $\F_{L,c}$ is equivalent to computing the rank of $\G_{L,c}$. To this end, define a nonzero vector ${\bf v} = [~{\bf v}_1^T~{\bf v}_2^T~]^T\in\R^{2N}$, which leads to
\begin{align}
    \G^T_{L,c}{\bf v}= \Re\{\A^H\text{diag}(\A\x)\}^T{\bf v}_1 + \Im\{\A^H\text{diag}(\A\x)\}^T{\bf v}_2.
\end{align}
Now let ${\bf u}={\bf v}_1 + j{\bf v}_2$, then
\begin{align}
    \G^T_{L,c}{\bf v} & = \Re\left\{ \left(\A^H\text{diag}(\A\x)\right)^H{\bf u} \right\} \notag\\
                  & = \Re\left\{ (\A\x)^*\odot(\A{\bf u}) \right\}
\end{align}
which equals to zero if and only if
\begin{align}
  \bfu = j\x
\end{align}
i.e.,
\begin{align}
    \G_{L,c}^T{\bf v} = \Re\left\{ j\left|\A\x\right|^2 \right\} = \0.
\end{align}
This means that there is only one direction ${\bf v} = [-\Im\{\x\}^T~\Re\{\x\}^T]^T$, which is non-zero, lies in the null space of $\G_{L,c}$, thus also in the null space of $\F_{L,c}$.

In the real $\x$ case, similar to the proof of $\F_{L,c}$, it suffices to show that there is no nonzero vector $\bfv\in\bR^N$ making $\F_{L,r}\bfv=\zero$. It is easy to see that
\begin{align}\label{Gr_zero}
  \G_{L,r}^T\bfv = \Re\left\{ (\A\x)^*\odot(\A\bfv) \right\}.
\end{align}
Since $\bfv$ is real-valued, it cannot be set to $\bfv=j\x$ to make \eqref{Gr_zero}  zero. Given a nontrivial $\A$, it is impossible to find a $\bfv$ such that $\A\bfv=\zero$ except ${\bf v}={\bf 0}$. Therefore, $\F_{L,r}$ is full rank. This completes the proof.


\begin{thebibliography}{1}

\bibitem{EUSIPCO2016}
C. Qian, X. Fu, N. D. Sidiropoulos, and L. Huang, ``Inexact alternating optimization for phase retrieval with outliers,'' \emph{Proceeding of 24th European Signal Processing Conference (EUSIPCO)}, pp. 1538-1542, Budapest, 2016.

\bibitem{GS}
R. Gerchberg and W. Saxton, ``A practical algorithm for the determination of phase from image and diffraction plane pictures,'' \emph{Optik}, vol. 35, pp. 237-246, 1972.

\bibitem{Fienup}
J. R. Fienup, ``Phase retrieval algorithms: A comparison,'' \emph{Applied Optics}, vol. 21, no. 15, pp. 2758-2769, 1982.

\bibitem{AltMinPr}
P. Netrapalli, P. Jain and S. Sanghavi, ``Phase retrieval using alternating minimization,'' \emph{IEEE Trans. Signal Process.}, vol. 63, no. 18, pp 4814-4826, 2015.

\bibitem{C1}
E. J. Cand\`es, T. Strohmer, and V. Voroninski. ``PhaseLift: Exact and stable signal recovery from magnitude measurements via convex programming,'' \emph{Communications on Pure and Applied Mathematics}, vol. 66, no. 8, pp. 1241-1274, 2013.

\bibitem{robustPL}
P. Hand, ``PhaseLift is robust to a constant fraction of arbitrary errors,'' \emph{Applied and Computational Harmonic Analysis}, vol. 42, no. 3, pp. 550-562, 2017.

\bibitem{WF}
E. J. Cand\`es, X. Li and M. Soltanolkotabi, ``Phase retrieval via Wirtinger Flow: Theory and algorithms," \emph{IEEE Trans. Info. Theory}, vol. 61, no. 4, pp. 1985-2007, 2015.

\bibitem{trunctedWF}
Y. Chen and E. Cand\`es, ``Solving random quadratic systems of equations is nearly as easy as solving linear systems,'' \emph{Advances in Neural Information Processing System}, pp. 739-747, 2015.

\bibitem{TAF}
G. Wang, G. B. Giannakis, and Y. C. Eldar, ``Solving systems of random quadratic equations via truncated amplitude flow,'' \emph{arXiv preprint arXiv:}1605.08285, 2016.

\bibitem{STAF}
G. Wang, G. B. Giannakis and J. Chen, ``Scalable solvers of random quadratic equations via stochastic truncated amplitude flow,'' \emph{IEEE Trans. Signal Process.}, vol. 65, no. 8, pp. 1961-1974, 2017.

\bibitem{FPP_PR}
C. Qian, N. D. Sidiropoulos, K. Huang, L. Huang and H. C. So, ``Phase retrieval using feasible point pursuit: Algorithms and Cram\'er-Rao bound,'' \emph{IEEE Trans. Signal Process.}, vol. 64, no. 20, pp. 5282-5296, 2016.

\bibitem{FPP_PR2}
C. Qian, N. D. Sidiropoulos, K. Huang, L. Huang and H. C. So, ``Least squares phase retrieval using feasible point pursuit,'' \textit{Proceedings of the International Conference on Acoustics, Speech, and Signal Processing (ICASSP 2016)}, Shanghai, China, pp. 4288-4292, 2016.


\bibitem{C2}
E. J. Cand\`es, Y. C. Eldar, T. Strohmer and V. Voroninski, ``Phase retrieval via matrix completion,'' \emph{SIAM Review}, vol. 57, no. 2 pp. 225-251, 2015.

\bibitem{IRLS_Quadratic}
J. Sigl, ``Nonlinear residual minimization by iteratively reweighted least squares,'' \emph{Computational Optimization and Applications}, vol. 64, no. 3, pp. 755-792, 2016.



\bibitem{PR_Spar_IN2}
D. S. Weller, A. Pnueli, G. Divon, O. Radzyner, Y. C. Eldar and J. A. Fessler, ``Undersampled phase retrieval with outliers,'' \emph{IEEE Trans. Computational Imaging}, vol. 1, no. 4, pp. 247-258, 2015.

\bibitem{ImpulsiveNoise}
S. Chen and H. Lu, ``Noise characteristic and its removal in digital radiographic system,'' \emph{Proceeding of 15th World Conference on Nondestructive Testing}, Roma 2000.

\bibitem{implusivenoise1}
I. Frosio and N. A. Borghese, ``Statistical based impulsive noise removal in digital radiography,'' \textit{IEEE Trans. Medical Imaging}, vol. 28, no. 1, pp.  3-16, 2009.

\bibitem{MTWF}
H. Zhang, Y. Chi and Y. Liang, ``Provable non-convex phase retrieval with outliers: Median truncated wirtinger flow,'' \textit{Proceeding of International Conference on Machine Learning (ICML)}, New York, 2016.





\bibitem{crb_pr0}
J. N. Cederquist and C. C. Wackerman, ``Phase-retrieval error: A lower bound,'' \emph{Journal of the Optical Society of America A}, vol. 4, no. 9 pp. 1788-1792, 1987.

\bibitem{crb_pr1}
R. Balan, ``The Fisher information matrix and the CRLB in a non-AWGN model for the phase retrieval problem,'' \textit{Proceeding of 2015 Internat. Conf. on Sampl. Theory and Applications (SampTA)}, pp. 178-182,  Washington, DC, 2015.


\bibitem{crb_pr3}
R. Balan, ``Reconstruction of signals from magnitudes of redundant representations: The complex case,'' \emph{Foundations of Computational Mathematics}, pp. 1-45, 2013.

\bibitem{crb_pr4}
A. S. Bandeira, J. Cahill, D. G. Mixon and A. A. Nelson, ``Saving phase: Injectivity and stability for phase retrieval,'' \emph{Applied and Computational Harmonic Analysis}, vol. 37, no. 1, pp. 106-125, 2014.



\bibitem{volmin}
X. Fu, K. Huang, B. Yang, W.-K. Ma and N.D. Sidiropoulos, ``Robust volume minimization-based matrix factorization for remote sensing and document clustering,'' \textit{IEEE Trans. Signal Process.}, vol. 64, no. 23, pp. 6254-6268, 2016.

\bibitem{outlier3}
Y. Liu, Y. Dai and S. Ma, ``Joint power and admission control: Non-convex $\ell_q$ approximation and an effective polynomial time deflation approach,'' \emph{IEEE Trans. Signal Process.}, vol. 63, no. 14, pp. 3641-3656, 2015.

\bibitem{outlier4}
R. Chartrand and W. Yin, ``Iteratively reweighted algorithms for compressive sensing,'' \emph{Proceeding of IEEE Internat. Conf. Acoust., Speech and Signal Process.}, Las Vegas, NV, pp. 3869-3872, 2008.


\bibitem{xf}
X. Fu, K. Huang, W.-K. Ma, N. D. Sidiropoulos, and R. Bro, ``Joint tensor factorization and outlying slab suppression with applications,'' \emph{IEEE Trans. Signal Process.}, vol. 63, no. 23, pp. 6315-6328, 2015.



\bibitem{beck2009fast}
A. Beck and M. Teboulle, ``A fast iterative shrinkage-thresholding algorithm for linear inverse problems,'' \textit{SIAM journal on Imaging Sciences}, vol.2, no.1, pp. 183--202, 2009.

\bibitem{cg}
Y. Dai and Y. Yuan, ``A nonlinear conjugate gradient method with a strong global convergence property,'' \emph{SIAM Journal on Optimization}, vol. 10, no. 1, pp. 177-182, 1999.

%
%
%
%
%


\bibitem{nesterov1}
Y. Nesterov, ``A method for unconstrained convex minimization problem with the rate of convergence O $(1/k^2)$,'' \emph{Doklady an SSSR}, vol. 269, no.3, pp. 543-547, 1983.

\bibitem{nesterov2}
O. Fercoq and P. Richt\`arik, ``Accelerated, parallel, and proximal coordinate descent,'' \emph{SIAM Journal on Optimization}, vol. 25, no. 4, pp. 1997-2023, 2015.

\bibitem{nesterov3}
Y. Nesterov, ``Efficiency of coordinate descent methods on huge-scale optimization problems,'' \emph{SIAM Journal on Optimization}, vol. 22, no. 2 pp. 341-362, 2012.


\bibitem{xuyin}
Y. Xu, R. Hao, W. Yin, and Z. Su, ``Parallel matrix factorization for low-rank tensor completion,'' \emph{Inverse Problems and Imaging}, vol. 9, no. 2, pp. 601-624, 2015.



\bibitem{crb1}
P. Stoica and T. L. Marzetta, ``Parameter estimation problems with singular information matrices,'' \emph{IEEE Trans. Signal Process.}, vol. 49, no. 1, pp. 87-90, 2001.

\bibitem{crb2}
A. O. Hero, III, J. A. Fessler and M. Usman, ``Exploring estimator bias-variance tradeoffs using the uniform CR bound,'' \emph{IEEE Trans. Signal Process.}, vol. 44, pp. 2026–-2041, 1996.


\bibitem{crb4}
K. Huang and N. D. Sidiropoulos, ``Putting nonnegative matrix factorization to the test: A tutorial derivation of pertinent Cram\'er-Rao bounds and performance benchmarking,'' \textit{IEEE Signal Processing Magazine, Special Issue on Source Separation and Applications}, vol. 31, no. 3, pp. 76-86, 2014.

\bibitem{LapNoiseCRB}
S. A. Vorobyov, Y. Rong, N. D. Sidiropoulos and A. B. Gershman, ``Robust iterative fitting of multilinear models,'' \emph{IEEE Trans. Signal Process.}, vol. 53, no. 8, pp. 2678-2689, 2005.

\bibitem{LapNoiseModel}
S. M. Kay, \emph{Fundamentals of statistical signal processing: Detection theory}, Upper Saddle River, NJ: Prentice-Hall, 1998.



\bibitem{basu2000}
S. Basu and Y. Bresler, ``The stability of nonlinear least squares problems and the Cramer-Rao bound,'' \emph{ IEEE Transactions on Signal Processing}, vol. 48, no. 12, pp. 3426-3436, 2000.



\bibitem{boundedness}
R. Chartrand and V. Staneva, ``Restricted isometry properties and nonconvex compressive sensing,'' \textit{Inverse Problems}, vol. 24, no. 3, pp. 035020, 2008.

\bibitem{bsum}
M. Razaviyayn, M. Hong, and Z.-Q. Luo, ``A unified convergence analysis of block successive minimization methods for nonsmooth optimization,'' \textit{SIAM Journal on Optimization}, vol. 23, no. 2, pp. 1126-1153, 2013.

\end{thebibliography}
\end{document}